# *Style and intensity of hydration among C-complex asteroids: a comparison to desiccated carbonaceous chondrites*

S. Potin[1], P. Beck[1], F. Usui[2], L. Bonal[1], P. Vernazza[3], B. Schmitt[1]

[1]Université Grenoble Alpes, CNRS, Institut de Planétologie et d'Astrophysique de Grenoble (IPAG), 414 rue de la Piscine 38400 Saint-Martin d'Hères, France. [2]Center for Planetary Science, Graduate School of Science, Kobe University, 7-1-48 Minatojima-Minamimachi, Chuo-Ku, Kobe 650-0047, Japan. [3]Laboratoire d'Astrophysique de Marseille (LAM), 38 rue Frédéric Joliot-Curie 13388 Marseille, France.

ABSTRACT

Here we report a comparison between reflectance spectroscopy of meteorites under asteroidal environment (high vacuum and temperature) and Main Belt and Near Earth Asteroids spectra. Focusing on the –OH absorption feature around 3µm, we show that the asteroidal environment induces a reduction of depth and width of the band, as well as a shift of the reflectance minimum. We then decompose the –OH feature into several components with a new model using Exponentially Modified Gaussians. Unlike previous studies, we confirm the link between these components, the aqueous alteration history and the amount of water molecules inside of the sample, using the shape of this spectral feature only. We then apply this deconvolution model to asteroids spectra which were obtained with a space-borne telescope and two space probes, and find a strong similarity with the components detected on meteorites, and among asteroids from a same type. Based on the conclusions drawn from our meteorites experiment, we suggest to use the 3-µm band as a tracer of the alteration history of the small bodies. Using the 3-µm band only, we show that Ryugu has been heavily altered by water, which is consistent with its parent body being covered with water ice, then went through a high temperature sequence, over 400°C. We also point out that the 3-µm band of Bennu shows signs of its newly discovered surface activity.

## 1. Introduction

C-complex asteroids represent around 60% of the mass of all known Main Belt Asteroids (MBAs) (DeMeo and Carry, 2013), taking Ceres into account for 35% of the mass of the belt (Vernazza and Beck, 2017). These objects are currently under high scrutiny by the planetary science community, and are often considered as primitive given their resemblance to laboratory spectra of carbonaceous chondrites. Amongst C-types asteroids, some objects show evidence of oxidized iron in the visible (Vilas, 1994) and the presence of an –OH related absorption feature in the 3-µm region (Rivkin et al., 2002). This fraction of C-type (Cg-, Ch-) appears to be strongly connected to CM chondrites (Fornasier et al., 2014; Andrew S. Rivkin et al., 2015; Vernazza et al., 2016) or some CR chondrites (Beck et al., 2018). In addition, a significant fraction of C-type asteroids do not show any absorption band in the visible and only a weak or absent 3-µm band. A first explanation to the lack of hydration signature is the occurrence of a thermal event on these objects leading to dehydration (Hiroi and Pieters, 1996; Hiroi and Zolensky, 1999). However, an alternative hypothesis is also proposed, that these objects were never hydrated, and are related to Interplanetary Dust Particles (IDPs) (Bradley et al., 1996; Vernazza and Beck, 2017).

When looking at objects from the Near-Earth Asteroids (NEAs) population, spectral signatures of hydration seems to be rare (Rivkin et al., 2015). Using Near Earth Objects delivery models, Rivkin and DeMeo (2019) calculated the theoretical amount of hydrated Ch-type asteroids among the NEAs population to be 17 ± 3 %, but observations revealed only 6 ± 3 % of Ch-type among NEAs. Two space missions are orbiting C-type NEAs at the time of this writing: the OSIRIS-REx space probe around (101955) Bennu (Lauretta et al., 2017) and the Hayabusa2 mission around (162173) Ryugu (Sugita et al., 2019). Interestingly, these two missions targets are strikingly different in their visible to near-infrared spectra: a clear 3-µm hydration feature seen by OSIRIS-REx on Bennu (Hamilton et al., 2019) and a tenuous feature seen by Hayabusa2 on Ryugu (Kitazato et al., 2019). These two missions might then be probing two different types of C-complex NEAs.

Thermal alteration was a major process among early accreted small bodies. It could have been induced by radioactive decay of $^{26}$Al (Keil, 2000; Neveu and Vernazza, 2019) and impacts (Bland et al., 2014). Finally, the solar irradiation, decreasing with increasing distance from the Sun, represents the main difference between MBAs and NEAs (Marchi et al., 2009). Because of their close distance to the Sun and chaotic trajectories, NEAs present a surface temperature higher than MBAs: between 150 K and 350 K around 1 au, and up to 1500 K during the closest possible approach to the Sun (Delbo and Michel, 2011; Dunn et al., 2013; Marchi et al., 2009).

In the present paper we focus on the comparison of hydration signatures between carbonaceous chondrites and C-type asteroids. In order to address the influence of asteroidal environment on the infrared signatures, meteorite samples were studied in laboratory under vacuum and with increasing temperature. We compare these laboratory data to those obtained on MBAs by the AKARI space telescope during the AKARI/IRC near-infrared asteroid spectroscopic survey: AcuA-spec (Usui et al., 2019), and on NEAs by the OSIRIS-REx and Hayabusa2 space probes. Finally, we discuss about the alteration history of the studied asteroids with a highlight on Ryugu and Bennu.

## 2. Samples and methods
### a. Selection of carbonaceous chondrites

The experiment was performed on a selection of 12 carbonaceous chondrites: 10 CM, 1 CR and 1 CI. The studied samples are presented in Table 1. Murchison and Orgueil have been obtained through museum loans (Beck et al., 2012a) and the others from the NASA collection (Meteorites Working Group).

| Meteorite name | classification and petrologic types |
|---|---|
| ALH 83100 | CM1/2 (1.1) |
| MET 01070 | CM1 (1.0) |
| DOM 08003 | CM2 (1.1) |
| Murchison | CM2 (1.6) |
| QUE 97990 | CM2 (1.7) |
| ALH 84033 | CM2 (heated) |
| EET 96029 | CM2 (heated) |
| MAC 88100 | CM2 (heated) |
| MIL 07700 | CM2 (heated) |
| WIS 91600 | CM2 (heated) |
| Orgueil | CI1 |
| GRO 95577 | CR1 |

Table 1: List of the studied carbonaceous chondrites and their petrographic grade. ALH = Allan Hills, DOM = Dominion Range, EET = Elephant Moraine, GRO = Grosvenor Mountains, MET = Meteorite Hill, MIL = Miller Range, QUE = Queen Alexandra Range, WIS = Wisconsin Range. Petrologic types from Rubin et al. (2007) and Alexander et al. (2013) under parenthesis.

b.  Reflectance spectroscopy under asteroid-like conditions

Reflectance spectroscopy was performed using the spectro-gonio radiometer SHADOWS (Spectrophotometer with cHanging Angles for the Detection Of Weak Signals) (Potin et al., 2018) at IPAG (Institut de Planétologie et d'Astrophysique de Grenoble, Grenoble, France). In this instrument, a monochromatic beam is focused on the sample and two detectors capture the reflected light. The spectra were acquired from 340 nm to 4200 nm. The spectral step was fixed at 10 nm from 340 to 1000nm, and from 2600 nm to 3700nm to have a better spectral sampling inside the absorption bands. The spectral step was then relaxed to 20 nm in the continuum between 1020 nm and 2600nm, and after 3720 nm. The spectral resolution of the goniometer varies over the whole spectrum, and is described in Table **2**. The illumination and emergence angles were set respectively to 0° and 30° from the normal of the surface.

| Spectral range | Spectral resolution |
|---|---|
| 340 nm – 679 nm | 4.86 nm – 4.75 nm |
| 680 nm – 1429 nm | 9.71 nm – 9.42 nm |
| 1430 nm – 2649 nm | 19.42 nm – 19.00 nm |
| 2650 nm – 4200 nm | 38.85 nm – 38.56 nm |

Table 2: Spectral resolution of the reflectance spectra.

The MIRAGE (Mesures en InfraRouge sous Atmosphère Gazeuse et Etuvée) environmental cell was developed to perform reflectance measurements with SHADOWS while simulating the high vacuum and warm environment of NEAs around the sample, from room temperature around 293K to 523K. MIRAGE consists of a circular vacuum chamber and a sample holder whose temperature is controlled by a 30 W heating resistor. Around 40 mg of material is needed to fill the sample holder and perform the experiment. The temperature is measured under the sample holder with a PT100 diode. A LakeShore™ Model 336 temperature controller provides the control and stabilization of the temperature inside the cell.

The top of the cell is closed by a sapphire window, enabling the light from the goniometer to be diffusely reflected by the sample, but also inducing parasitic reflections between the sample and the sapphire window itself (Pommerol et al., 2009). The photometric effects of these reflections have been calibrated and are removed during data reduction.

MIRAGE is connected to a high vacuum pump maintaining a pressure between $10^{-7}$ and $10^{-6}$ mbar. It is possible to inject a few millibars of gaseous nitrogen inside the cell to provide a better thermalization of the sample, but the choice was made to keep working under dynamic vacuum because of the targeted temperature up to 523 K.

     c.     Sample preparation and measurement protocol

The samples were first manually powdered into an agate mortar but not sieved to keep a large distribution of grain sizes. The sample holder was then filled with the obtained powder and installed in the chamber. The experiment started when the pressure inside the cell was lower than $10^{-4}$ mbar to remove the contribution of the possibly adsorbed terrestrial water. Spectra were acquired around room temperature 297 K, 373 K to ensure the removal of all adsorbed atmospheric molecules in the sample, the highest reachable temperature 523 K, and again at 297 K to check for any potential, macroscopic physico-chemical changes in the sample. For each thermal step, the sample was kept at the desired temperature during 30 minutes for the sample to thermalize, then the reflectance acquisition started for 1 hour. The temperature of the sample was monitored during the whole acquisition. Variations of less than 1 K were observed when at room temperature, and these variations weaken down to 0.005 K at 523 K.

     d.     Deconvolution of the 3-µm band

The 3-µm band corresponds to vibration modes of –OH groups. Bishop et al. (1994) and Frost et al. (2000) presented a description of the 3-µm band as a convolution of 3 components:

     - vibrations of –OH groups typical of hydrated minerals around 2.7-2.8 µm (hereafter called metal-OH component). Weakly bonded interlayer water can also provide a tenuous contribution to this component around 2.75 µm (Kuligiewicz et al., 2015).

     - symmetric and asymmetric stretching vibrations around 3.1 µm of 'bulk' $H_2O$ molecules strongly bound to cations within the mineral, crystal structure or inside clay layers (hereafter called low frequency water LFW, or structural water component)

- symmetric and asymmetric stretching vibrations around 2.9 μm of $H_2O$ molecules adsorbed on the surface of the grains, or trapped into mesopores (hereafter called high frequency water HFW, or adsorbed water component)

In order to study the different contributions of the 3-μm band in all spectra and their evolution with temperature, the complete absorption feature was decomposed into two or three components. The description of the model, the error calculations and the comparison with existing models are described in Potin et al.(2020).

## 3. Results

The reflectance spectra acquired at room temperature before and after the heating experiment are presented in Figure **1**. The reflectance ratio between the last spectra acquired after the heating and the first is also shown for each sample.

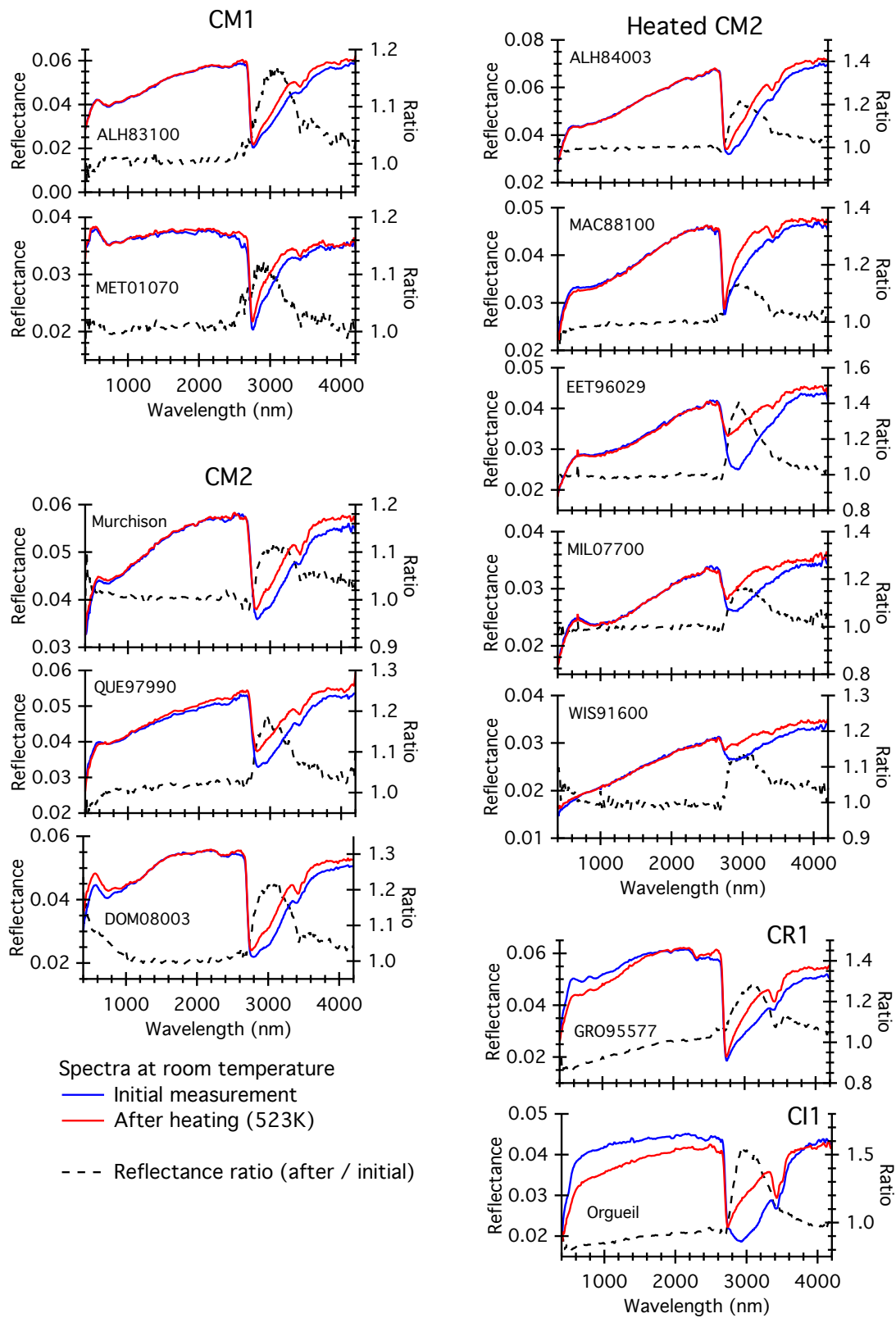

Figure 1: Reflectance spectra acquired at room temperature, before and after heating experiment, of the selected chondrites. Blue: spectra at the beginning of the heating experiment (297K). Red: last spectra of the experiment after heating at 523 K once the sample has cooled down to 297K. Black: reflectance ratio between the last (after heating) and initial (before heating) spectra.

a. The effect of heating on the hydration 3-µm feature

We use four parameters to analyze the global shape of the 3-µm band: the amplitude, the broadness (represented by the Full Width at Half Maximum FWHM), the position of the minimum of reflectance and a symmetry factor.

The amplitude of the band, or band depth, is calculated at the minimum of reflectance $\lambda_{center}$ assuming a linear continuum between 2600 and 4000 nm (Clark and Roush, 1984):

$$Amplitude = 1 - \frac{Refl_{\lambda_{center}}}{Continuum_{\lambda_{center}}}$$

with $Refl_{\lambda_{center}}$ the value of reflectance and $Continuum_{\lambda_{center}}$ the calculated value of the continuum at the position $\lambda_{center}$.

The symmetry factor is calculated as

$$Sym = \frac{\lambda_{center} - \lambda_A}{\lambda_B - \lambda_{center}}$$

with $\lambda_{center}$ the position of the band minimum and $\lambda_A$ and $\lambda_B$ the position of the band taken at half maximum. A symmetry factor of 0 characterizes a sharp left wing of the band while a symmetry factor of 1 represents a perfectly symmetrical band.

The band parameters before and after heating at 523K are presented in Figure **2**. All calculated parameters can be found in the Appendix.

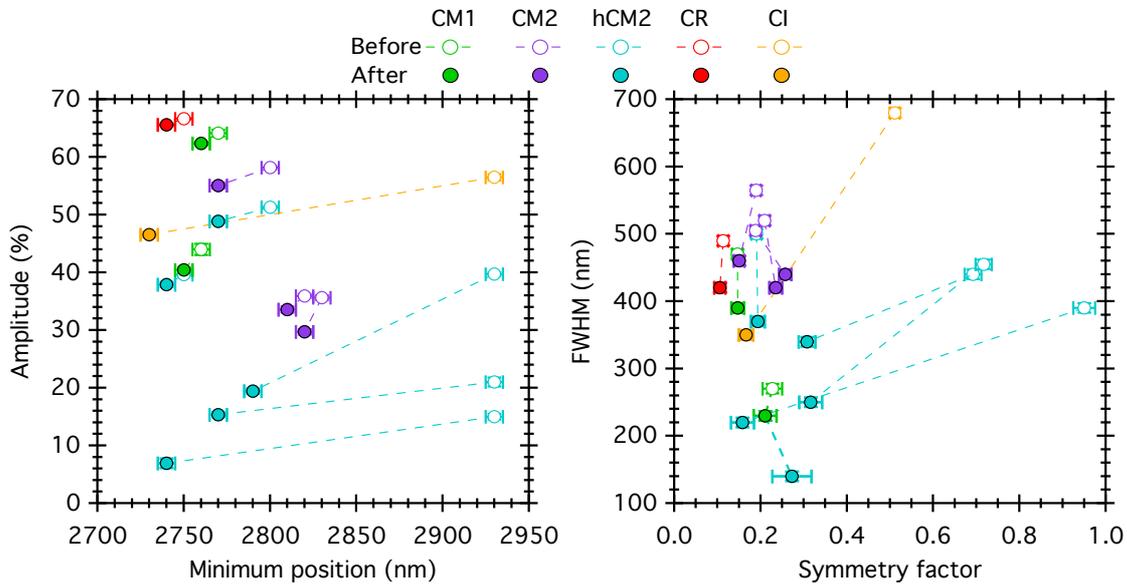

Figure 2: Variation of the amplitude, position, FWHM and symmetry factor of the 3-µm band before and after heating at 523K. Open circles: parameters before heating. Filled circles: parameters after heating.

In all spectra, we report a decrease of the amplitude and FWHM of the 3-µm band, along with a shift of the minimum position toward short wavelengths. The two studied CM1 and the CR chondrites show the least variations (a loss of less than 2 % in amplitude and a shift of 10 nm), while the heated CM2 present drastic variations during the temperature experiment: a wavelength shift up to 205 nm, a loss of amplitude up to 20 % and of width up to 330 nm.

Phyllosilicates are the result of the aqueous alteration of silicates (Beck et al., 2014; Brearley, 2006; Ikeda and Prinz, 1993; Keil, 2000) which are a major constituent of heavily altered chondrites. Phyllosilicates in CM, CR and CI chondrites are de-hydroxylated by temperatures over 680 K (Garenne et al., 2014; King et al., 2015) and so cannot be altered by our heating experiment. Therefore, the heavily altered chondrites mainly composed of phyllosilicates will present only weak variations during the heating. Other components such as adsorbed molecular water, mesopore water, interlayer or structural water, however, can be altered by temperatures lower than 523K and thus their evolution is detected during our experiment.

We now separate the different components of the 3-µm band, as well as the organic features. The result is displayed in Figure **3** for Orgueil. Though we expect 4 or 5 components to contribute to the organic bands between 3.35-3.50 µm, corresponding to the various stretching modes in CH, $CH_2$ and $CH_3$ groups (Orthous-Daunay et al., 2013), only two are modeled here, as they are not all resolved at our spectral resolution. The decomposed spectra of all studied meteorite samples are available in the Appendix.

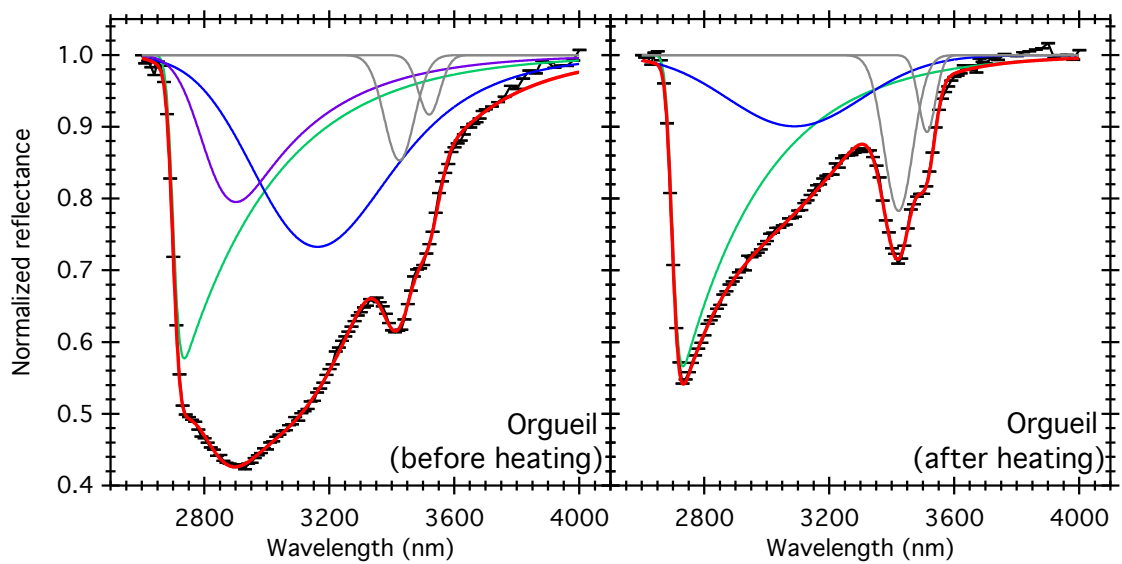

Figure 3: Example of mathematical decomposition of the components of the 3-µm band in the case of the CI chondrite Orgueil before (left) and after (right) heating at 523K. Black dots: measurement data; Red: modeled spectrum; Green: hydrated minerals component (metal-OH); Purple: HFW component (adsorbed water); Blue: LFW component (structural and interlayer water); Grey: organic features.

The metal–OH phyllosilicates component in Figure **3** peaking at 2740 nm is present in the spectra of each sample, before and after the heating sequence. The reduction of the LFW (structural and interlayer water) and disappearance of the HFW (adsorbed water) components by the increasing temperature leaves the hydroxyl feature as a major contributor to the band. The minimum of reflectance shifts from 2930 nm towards 2750 nm as shown in Figure **2** and the round band changes to a sharp triangular shape. All components of the 3-µm band are affected, at varying degrees, by the high temperature.

The different parameters (band depth, minimum position, FWHM and symmetry factor) for all meteoritic samples, before and after the heating experiment, are available in the Appendix. The variations of these parameters during the heating experiment are presented in Figure **4**.

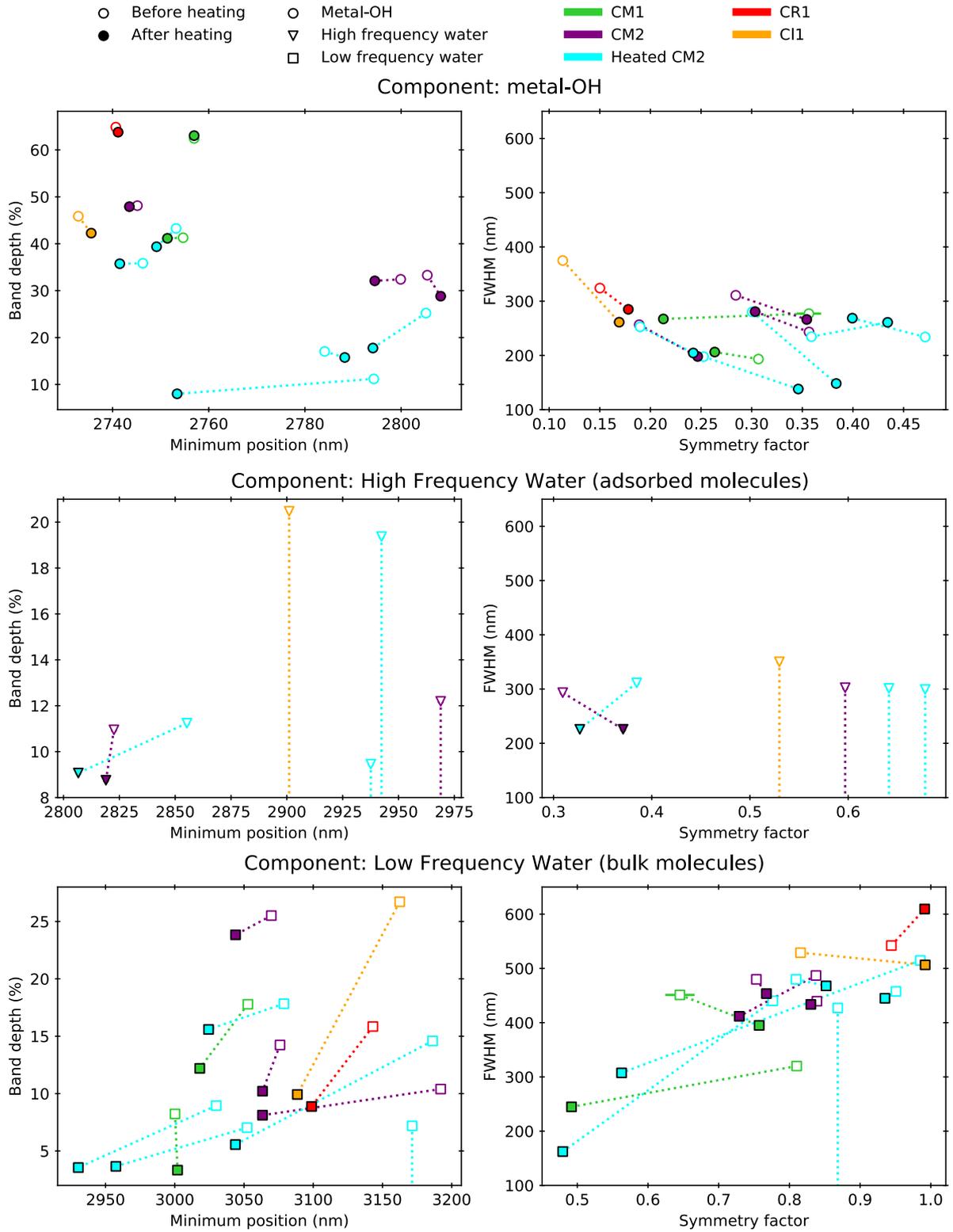

Figure 4: Variations of the components of the 3-µm band during heating up to 523K. Top panels (circles): component due to the metal-OH, Middle panels (triangles): component of high frequency water molecules, Bottom panels (squares): component of low frequency water molecules. Open markers: component parameters before heating at 523K, Filled markers: component parameters after heating. Components disappearing due to the temperature are noted with a dotted line without a filled marker at the end.

Before presenting the effects of the increasing temperature, it is important to note that the initial parameters of the components directly depend on the hydration state of the sample. With increasing aqueous alteration, the metal-OH component deepens and shifts toward the short wavelengths. Aqueous alteration induces the formation of Mg-bearing phyllosilicates, which are characterized by an absorption feature around 2.71 µm in carbonaceous chondrites (Beck et al., 2010). The most aqueously altered chondrites will present a first component around 2.7 µm, while it will be detected at longer wavelengths for the least altered meteorites.

The second and third components, respectively due to the HFW and LFW molecules, depend on the number of water molecules inside the sample. An increase in the concentration of $H_2O$ molecules in the sample will increase the H-bonding between water molecules, so inducing a lowering of the stretching frequency of the OH bonds (Kuligiewicz et al., 2015; Schultz, 1957). This lowered vibration frequency induces a shift of the absorption band toward longer wavelengths. The increase of amplitude and width of the bulk water and adsorbed component is a direct effect of the increase of the number of water molecules in the sample (Bertie et al., 1989). Half of the samples (ALH 83100 (CM1), GRO 95577 (CR1), MAC 88100 (heated CM2), MET 01070 (CM1), Murchison (CM2), QUE 97990 (CM2) and WIS 91600 (heated CM2)) do not show the second component (HFW) before heating, indicating that the water is part of the mineral structure rather than adsorbed on the grains. Four out of six others samples have lost this component after heating.

The broadness and symmetry factor of the metal-OH component are significantly affected by the heating experiment. As phyllosilicates are not dehydroxylated at 523 K, the position and amplitude of this component are not significantly altered. However, the thinning of the band of half of the samples could be explained by the dehydration of the low amount, or traces, of others –OH-bearing minerals, or by the removal of the small 2.75 µm contribution of interlayer water. With increasing temperature, the amount of water molecules, bound to cations and adsorbed, decreases until full dehydration. Most of the adsorbed water is removed from the sample, as can be seen by the disappearance of the HFW component on most samples. With increasing temperature, the LFW component shifts toward the short wavelengths. This can be explained by a decrease of the abundance of water molecules inside the sample, also explaining the loss of the amplitude and width, which reduces the H-bonding between the $H_2O$ molecules (Schultz, 1957).

b.     Reflectance ratio

We calculated the ratio of two separated reflectance values, as the method used in Usui et al. (2019):

$$Ratio = \frac{R_{2.45}}{R_{0.55}}$$

with $R_{2.45}$ and $R_{0.55}$ respectively the reflectance measured at 2.45 µm and 0.55 µm. The modification of the reflectance ratio during the heating experiment is presented in Figure **5**.

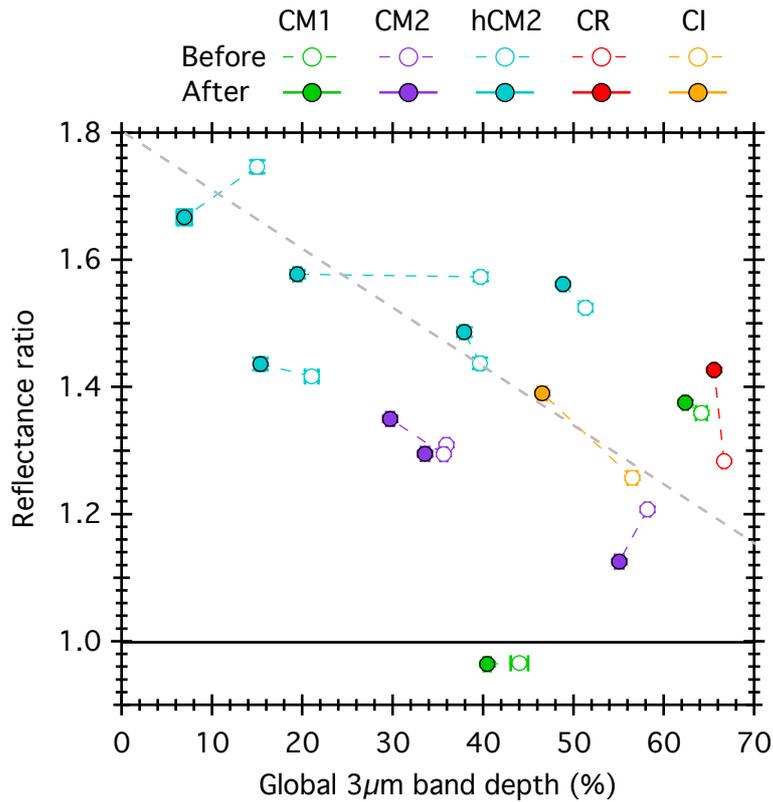

Figure 5: Variation of the reflectance ratio compared to the global 3-µm band depth during the heating experiment. The black line marking a ratio of 1 separates the blue spectra (decreasing reflectance with increasing wavelength) from red spectra (increasing reflectance with increasing wavelength). The dotted grey line indicates the negative trend between the reflectance ratio and band depth.

In Figure **5**, one can notice that the heated carbonaceous chondrites tend to increase their reflectance ratio (i.e. to become "redder") compared to the corresponding unheated samples. Regarding the effect of temperature, no general trend is seen in Figure **5**, as 6 out of 10 chondrites show an increase of the ratio, 2 samples are characterized by a decrease and 4 by no obvious change, independently of the petrographic grade.

In Figure **5**, it is interesting to note the negative trend between the ratio of the reflectance spectrum and the depth of the 3-µm band. Based on heating experiments performed on the Murchison meteorite, such a correlation is not expected. Indeed, heated Murchison samples show only minor changes in slope, except at the highest temperature (>1000 K) where the sample becomes bluer (Hiroi et al., 1996). However, in our study, the slope of heated CM chondrites appears to be higher in general than non-heated CM-chondrites. Such a slope change could be related to a progressive change in the structure of the macromolecular carbonaceous compounds (Kaluna et al., 2017; Lantz et al., 2015).

c. Alteration of the carbonaceous matter

The reflectance spectra measured after heating reveal an increase of the depth and width of the organics features when exposed to high temperature. It is important to note that these organic features are easily detectable in all spectra after the heating sequence, even if the initial measurement show no or weak signs of the 3.42-3.51 µm bands. The organics features are located on the right wing of the 3-µm deep hydration band, which gets thinner during our heating experiment. The partially or fully hidden organic features on the initial spectra can then be revealed, which can induce an apparent increase of the bands. However, the spectral modeling of the complete feature, -OH and –$CH_2$/–$CH_3$ combined (Figure **3**), shows a clear increase of the depth of the organic bands. In the case of Orgueil, the organic component at 3420 nm deepens from 16.3 % to 21.6 %, and the other at 3510 nm from 9 % to 11.5 %.

### 4. Comparison with asteroids and discussion

The observations with the AKARI telescope and the associated data processing are described in Usui et al. (2019), where the reflectance spectra of 64 MBAs are provided[1]. With up to 12 parameters, the model decomposing the 3-µm band can have some difficulties to converge on noisy spectra. To avoid misinterpretation of spectral features, we selected only those spectra showing a well-defined 3-µm band, with small error bars along the short wavelength wing. The 11 selected spectra are shown in Figure **6**. The reflectance spectra of the NEAs Ryugu and Bennu have been digitalized respectively from Kitazato et al. (2019) and Hamilton et al. (2019), and are presented in Figure **7**.

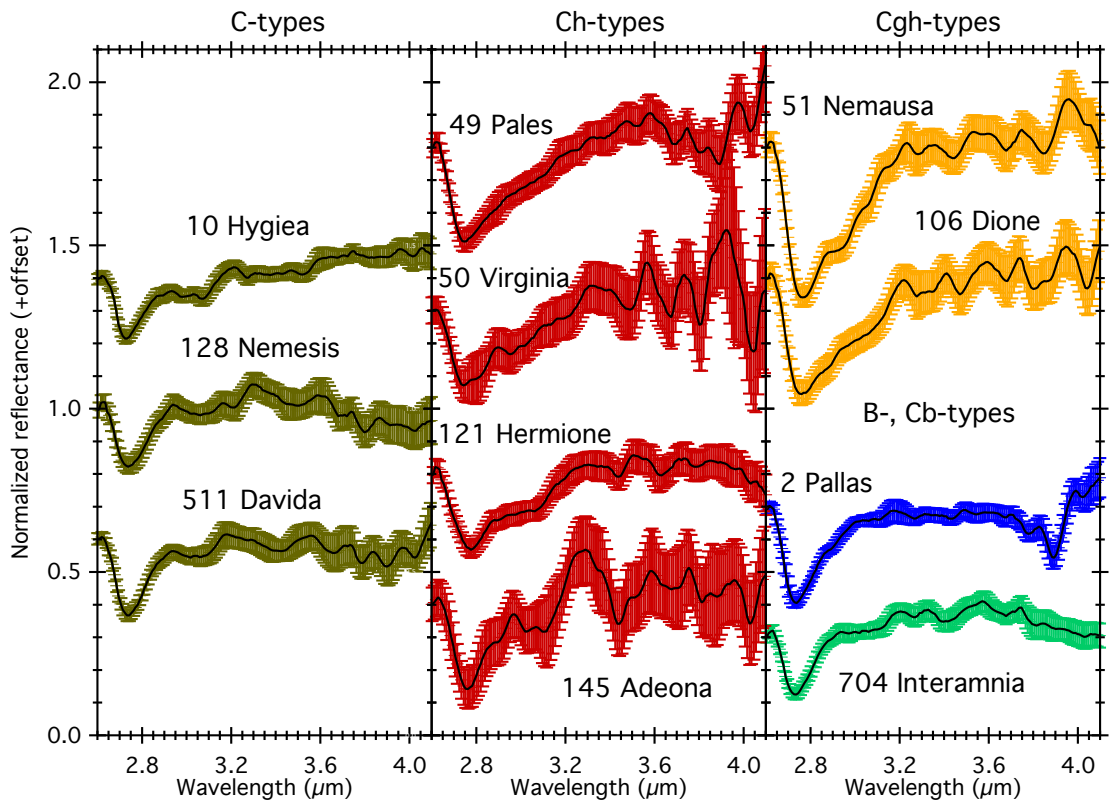

Figure 6: Selected AKARI spectra of C-complex asteroids. Black: reflectance observations, colors: error bars. The colors given to the different asteroid types will be similar in the following figures. The spectra have been normalized to 2.55 µm.

---

[1] The corrected spectra of the AKARI data base are available to the public on the JAXA archive: http://www.ir.isas.jaxa.jp/AKARI/Archive/

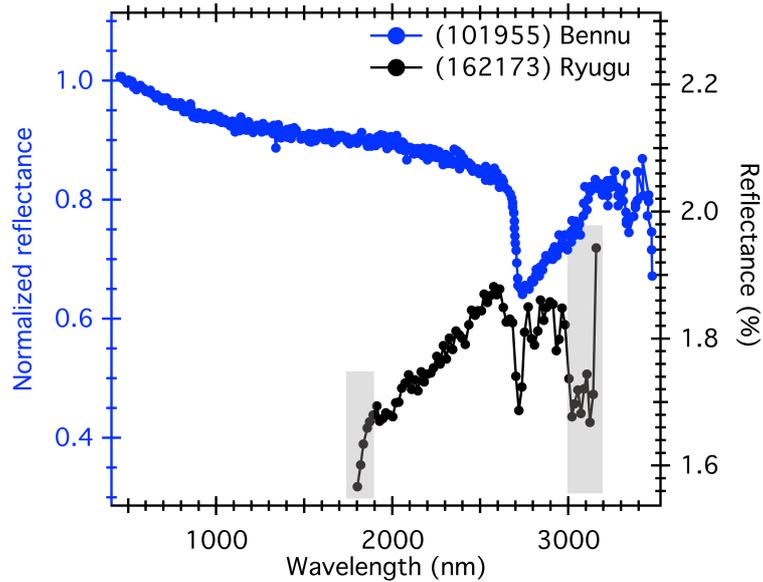

Figure 7: Reflectance spectra of the NEA Ryugu (absolute reflectance values) and Bennu (reflectance normalized at 550 nm). The spectra have been digitalized from Hamilton et al (2019) for Bennu and Kitazato et al. (2019) for Ryugu. The shaded area on the Ryugu spectrum indicate the region with large calibration residuals, as described in Kitazato et al. (2019).

The disk integrated measurements from AKARI were obtained at solar phase angle between 16° and 25° (Usui et al., 2019) and Bennu at approximately 5.2° (Hamilton et al., 2019) also integrated over the whole surface of the small body. The spectrum of Ryugu used in this study was also acquired near opposition but during an equatorial scan and thus covering a small part of the surface of the asteroid (Kitazato et al., 2019).

In this section, we will compare spectra from AKARI, OSIRIS-REx and Hayabusa2 to our meteorites measured in asteroid-like conditions. For clarity, only the parameters calculated on the spectra measured after the heating (i.e. under asteroid-like conditions) are displayed on the following figures.

a. Shape of the 3-µm band

In a first step, the global parameters of the complete 3-µm band (amplitude, position, FWHM and symmetry factor) of the asteroids spectra are calculated using the same method as described earlier for the meteorite spectra. Figure **8** compares the shape of the 3-µm band on the asteroids and meteorites spectra (after heating at 523 K).

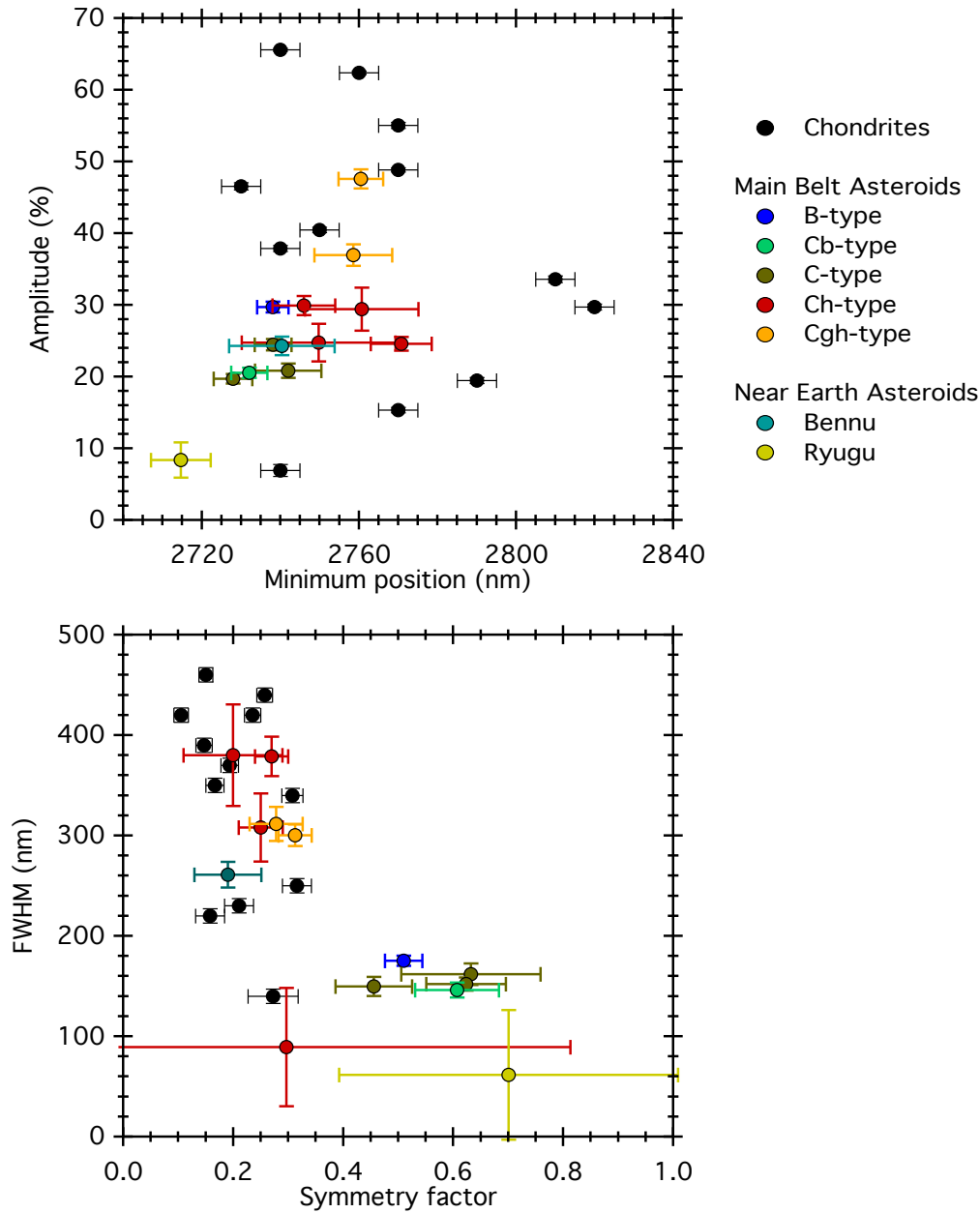

Figure 8: Comparison between the shape of the complete 3-µm band of the asteroids (MBAs and NEAs) spectra (colored circles) and chondrites after the heating experiment (black circles).

Previous analyses pointed out the strong dependence of the shape of the reflectance spectra with the observation geometry (Pommerol and Schmitt, 2008; Potin et al., 2019; Schröder et al., 2014). It has been shown that the amplitude of the detected absorption band decreases with increasing phase angle. No effects on the width and position of the bands have been previously detected. As the AKARI spectra were acquired with a phase angle similar to our experiment (30°) and the NEAs near opposition, we do not expect a noticeable effect of the geometry when comparing asteroids and meteorites spectra. However, the texture of the surface itself also plays a role on the detected absorption bands, where a macroscopic chip of meteorite will present weaker features than the same samples crushed into a powder (Potin et al.,

2019). Our experiment uses powdered samples with a large distribution of grains sizes, which is not necessarily representative of the surface texture of asteroids with fine and/or coarse regolith and boulders.

The minimum positions of the bands on the asteroidal spectra are consistent with most of our measurements on carbonaceous chondrites after heating. However, only the amplitudes of the bands on both Cgh-type asteroids are also consistent with these measurements. The Ch-, C-, Cb- and B-types, as well as both NEAs show weaker 3-µm absorption band compared to 7 of our meteorites spectra. The other 5 meteorite spectra either display a strongly shifted band or a much weaker absorption, but none have both the position and amplitude of most of the observed spectra. This could be an effect of the difference of texture between our powdered meteorites and the surface of the asteroids. The FWHM and symmetry factor of the Ch- types (except Adeona), Cgh- types and Bennu correspond to the values derived from our meteorites spectra. But the C-, B-, Cb-types, Adeona and Ryugu show thinner and more symmetrical absorption features compared to our meteorites. Though the texture of the sample could induce variations in the depth of the 3-µm band, its impact on the FWHM and symmetry of the feature is yet to be determined. Moreover, Matsuoka et al. (2015) analyzed the effect of space weathering on the reflectance spectra of Murchison and found a decrease of amplitude of the 3-µm band of nearly 35% with increasing irradiation. The difference of amplitude between the asteroidal spectra and some of our measurements could thus be due to the space weathering occurring on the surface of the small bodies, compared to our unaltered meteorites.

We then compare the decomposition of the 3-µm band of asteroids spectra with our measurements on meteorites. All the modeled spectra are presented in the Appendix.

All the meteorites spectra can be modeled using two well-defined components: the metal-OH and a water component. For two asteroids spectra, this second component is absent, leaving only the metal-OH components to appear as it is the case for Pallas and Ryugu.

We hereafter present an investigation of the alteration history of C-complex asteroids based on the spectral variations of the 3-µm band components. Figure **9** presents the parameters (band depth, FWHM, minimum position and symmetry factor) of the metal-OH and water component derived from the asteroids and meteorites spectra (after heating at 523 K).

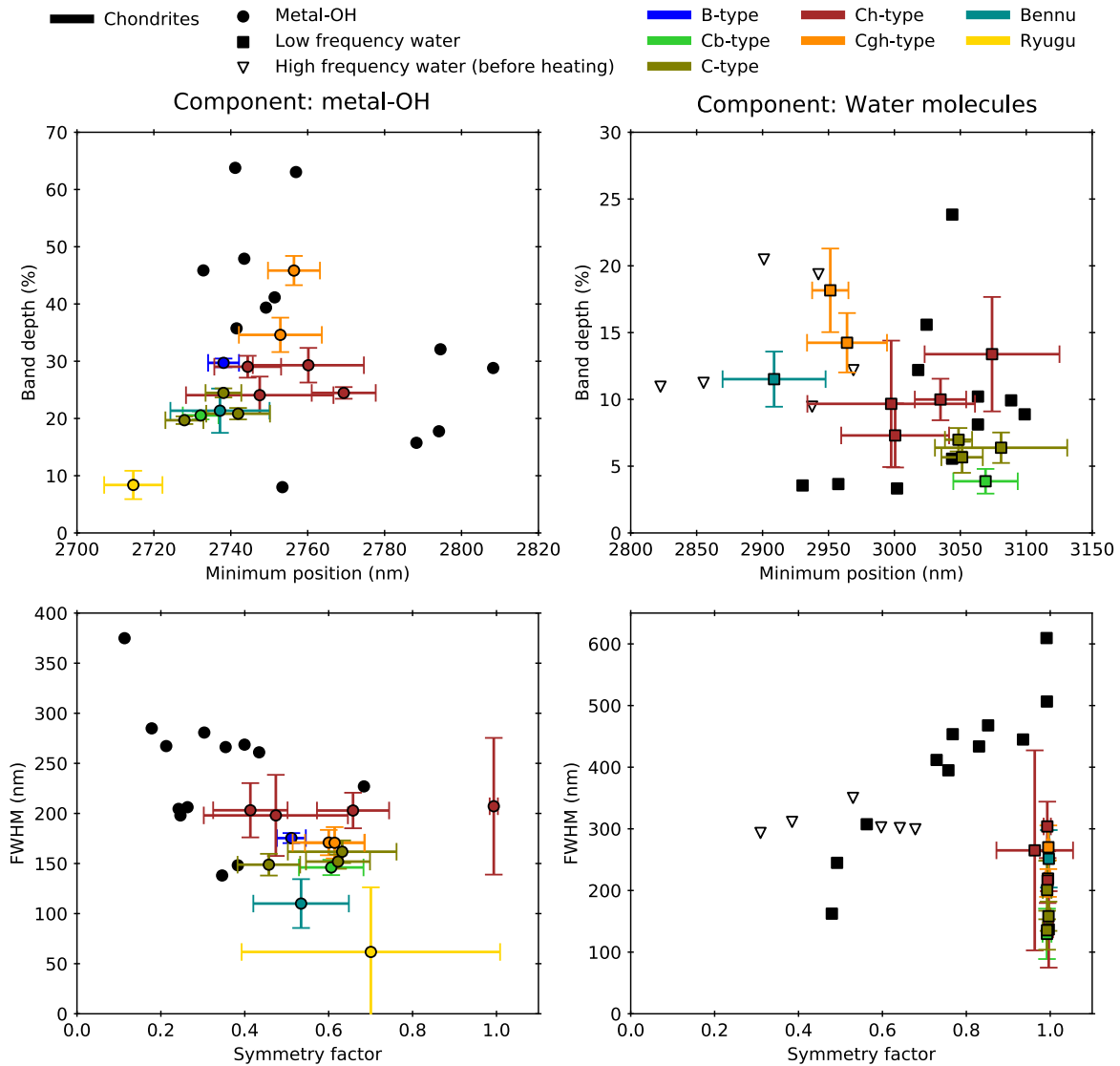

Figure 9: Comparison of the parameters of the metal-OH (left) and LFW / HWF (right) components of the 3-μm band between asteroids observations (colored symbols) and meteorites spectra (this study, in black).

First, it is important to note that asteroids from the same type present roughly the same components, sign of a similar alteration history among them.

As for the global 3 μm band, the metal-OH components detected on asteroids tend to be weaker than the ones detected on meteorites, but the positions of their minimum match the values found on most of the meteorites. The metal-OH components of asteroids are also thinner and more symmetrical than for meteorites, but as it can be seen in Figure **9**, this component follows the same trend of increasing symmetry with decreasing width as seen on our series of meteorites. Both the amplitude and position of the water component of asteroids are consistent with the studied carbonaceous chondrites. However, like the metal-OH component, the water feature is thinner on the asteroids spectra. This water component on asteroids seems perfectly symmetrical while meteorites show a correlation between its symmetry and width. However, this difference is most probably due to the large error bars in the asteroids spectra around this component, lowering the constraints on the asymmetry parameter in the fit model.

The different effects of aqueous and thermal alteration highlighted by the meteorites experiment can then be applied to the asteroids.

- (2) Pallas and (704) Interamnia are classified as B-type and Cb-type, respectively. They present a well-defined metal-OH component from hydrated minerals, centered at 2738 ± 4 nm for Pallas and 2732 ± 4 for Interamnia. This is consistent with the component detected on our most aqueously altered chondrites: the CR1 GRO95577 (centered at 2741.15 ± 0.05 nm) and the CI1 Orgueil (centered at 2731.76 ± 0.02 nm). This reflects a heavy alteration by water on the surface of the asteroids, though the absence of detection of structural water molecules on the spectrum of Pallas suggests a strong dehydration by exogenic process as was proposed by Schmidt and Castillo-Rogez (2012). Interamnia presents the shallowest water component of the asteroids group, with a band depth of 3.8 ± 0.9 % and centered at 3069 ± 24 nm. As seen from the heating experiment, a shift of the LFW component toward longer wavelengths suggests an increase of the number of structural water molecules inside the minerals, but this is incoherent with the small amplitude of the component. This could be the effect of a different surface composition, and thus mineralogical structure on Interamnia, gathering the water molecules together and shifting the position of this water component.

- The components detected on the spectra of the C-types (10) Hygiea, (128) Nemesis and (511) Davida are similar to those detected on (2) Pallas and (704) Interamnia. Their metal-OH components are centered between 2728 ± 5 nm for Hygiea and 2742 ± 8 nm for Nemesis. This traces a strong alteration of the surface, stronger than the studied Ch- and Cgh-types, and similar to what is observed on the CR1 GRO95577 and the CI1 Orgueil. The three C-types present a second component due to LFW molecules, slightly deeper than the component detected on Interamnia.

- (49) Pales, (50) Virginia, (121) Hermione and (145) Adeona are classified as Ch-type asteroids. Their metal-OH components are centered from 2744 ± 9 nm for Pales to 2769 ± 8 nm for Hermione. This is consistent with type 1 and 2 carbonaceous chondrites, like the CR1 GRO95577 (centered at 2741 nm) and the CM2 Murchison (centered at 2794.53 ± 0.03 nm). All studied Ch-type asteroids present a second component due to structural water molecules. The minimum position and amplitude of their components are similar to those typical of LFW molecules observed on meteorites spectra. The general trend of increasing amplitude with shift towards longer wavelengths is also observed. We can suggest an increasing amount of structural water molecules in the studied Ch-types, starting from Pales (amplitude of 7.3 ± 2.4 % at 3000 ± 40 nm) to Adeona (amplitude of 13 .4 ± 4.3 % at 3074 ± 51 nm).

- The metal-OH components are detected on the Cgh-type asteroids (51) Nemausa and (106) Dione respectively at 2756 ± 7 nm and 2753 ± 10 nm. This falls between the two studied CM1 chondrites ALH 83100 (component centered at 2756.96 ± 0.03 nm) and MET 01070 (component centered at 2751.01 ± 0.03 nm). Both asteroids present the deepest metal-OH component of the asteroid group, with an amplitude of 45 ± 2 % for Nemausa and 34 ±3 % for Dione. Their reflectance spectra also present a water component, centered respectively at 2951 ±13 nm and 2964 ± 30 nm, with an amplitude of 18 ± 3 % and 14 ± 2 %. These second components are similar in terms of amplitude and minimum position to the component due to HFW (adsorbed water) detected on meteorites spectra. Taking into account the extremely low pressure at the surface of the small bodies, the detection of adsorbed water molecules is surprising. Moreover, this is at odd with what has been observed during our experiment with meteorites where the LFW is always observed, and the component due to adsorbed water frequently disappearing after heating under vacuum. As Dione was proposed as an active asteroid (Clark et al., 2008), we can make the hypothesis that the position of this band corresponds to the detection of adsorbed water molecules on the surface of the body as a consequence of this activity. Given the position and band depth of the component, we can also suggest a possible surface activity on Nemausa.

- The reflectance spectrum of Bennu presents a metal-OH component at 2737 ± 12 nm that traces a strong aqueous alteration of the surface. A second component due to water molecules is detected on the reflectance spectrum of Bennu (Figure **7**). Its minimum position and amplitude is similar to the HFW component derived from meteorites measurements. This could indicate the presence of adsorbed water molecules on the surface of the asteroid, a consequence of the newly discovered activity of the surface (Connolly et al., 2019).

- Ryugu is the most aqueously altered object of this analysis, with its hydrated component centered at 2714 ± 7 nm. This is consistent with the suggestion that Ryugu has been produced from a parent body containing water ice (Sugita et al., 2019). If we consider Ryugu similar to CM or CI chondrites (Kitazato et al., 2019; Sugita et al., 2019), this band position corresponds to the one observed for $Mg_3OH$ in saponite (Kuligiewicz et al., 2015), and thus suggests a surface composition almost exclusively made of phyllosilicates. However, the small amplitude of this component, 8.3 ± 2.5 % and the absence of detected water molecules are in favor of an important thermal alteration over 680 K. At this temperature, all water molecules and hydroxyls in the surface minerals, however strongly bound, are released. The phyllosilicates are desiccated, reducing the metal-OH component to barely detectable. Analyses of the boulders, impact craters and regolith on the surface of Ryugu by Sugita et al., (2019) suggested that this strong thermal alteration could have occurred on an older and larger parent body, before the catastrophic disruption forming Ryugu.

     b.     Reflectance ratio (R2.45/R0.55)

Figure **10** compares the reflectance ratio of all the studied objects, meteorites and asteroids. Values of the ratio of the reflectance spectra of asteroids are taken from Usui et al. (2019). The value for Bennu has been calculated using the same reflectance ratio as before on the digitalized spectrum from Hamilton et al. (2019). As the Ryugu spectrum published by Kitazato et al. (2019) covers only the infrared range, the reflectance ratio of this small body has not been calculated.

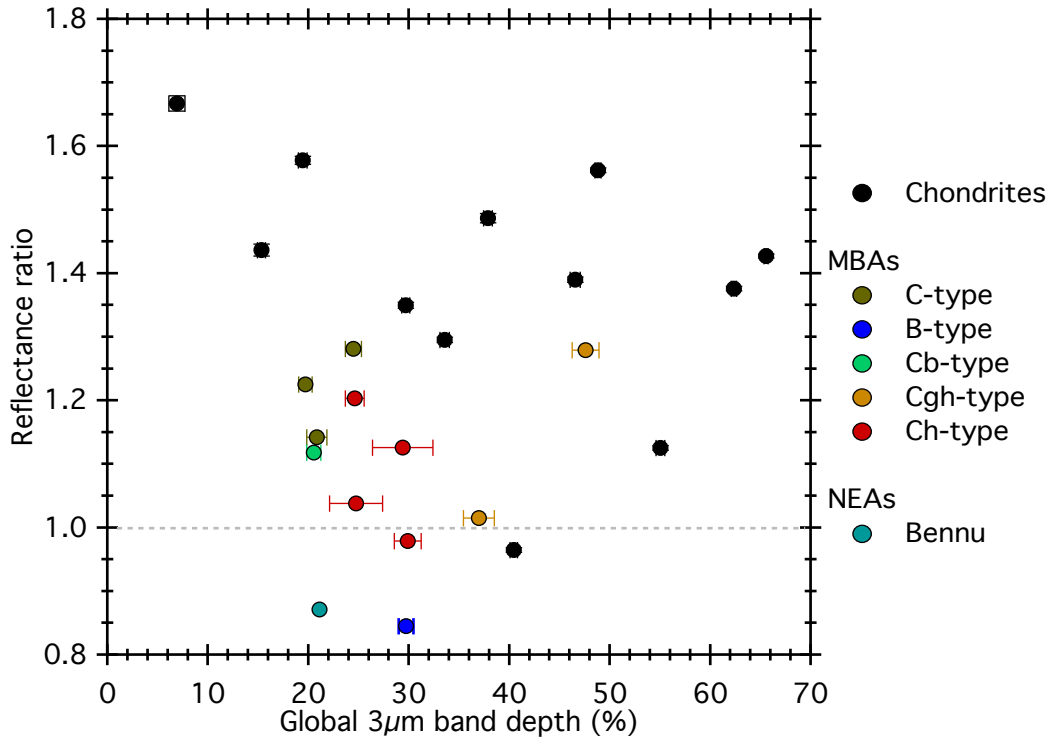

Figure 10: Comparison of the reflectance ratio and 3-μm band depth from asteroids (AKARI and OSIRIS-REx data) and meteorites (present work, after heating at 523 K). MBAs ratio from Usui et al. (2019), the ratio value for Bennu has been calculated from the digitalized spectrum. No data is available for Ryugu. The dotted grey line marking a ratio of 1 separates the blue spectra (decreasing reflectance with increasing wavelength) from red spectra (increasing reflectance with increasing wavelength).

For the same amplitude of the 3-μm band, the asteroids show bluer spectra (i.e., lower ratio) compared to the meteorite spectra. The average of all ratio values calculated on the meteorites spectra reaches 1.35±0.24, for 1.09±0.15 on the asteroid spectra. Three effects can lead to this difference of less than 25%:

- Observation geometry has a strong effect on the slope of the reflectance spectra. The strengthening of the spectral slope with increasing phase angle, called phase reddening, has been observed on asteroids (Binzel et al., 2015; Blanco and Catalano, 1979; Clark et al., 2002; Gehrels et al., 1970; Sanchez et al., 2012) and in laboratory on meteorites (Beck et al., 2012b; Potin et al., 2019; Schröder et al., 2014). In this study, all our measurements are made with a 30° phase angle, while the observations in Usui et al. (2019) are measured with phase angle varying for each asteroid, all of them around 20°. A difference of 10° in the phase angle between the observations and our measurements is too small to explain the redder slope seen in our meteorites.

- It has been shown during laboratory measurements that reflectance spectra of compacted surface appear bluer than those obtained on powders (Britt and Pieters, 1988; Cloutis et al., 2011; Potin et al., 2019). Binzel et al., (2015) found a variation of around 16% of the spectral slope value due to varying grain size of the sample. Our samples were manually ground into a powder to simulate the regolith-covered surface of Main Belt Asteroids, thus the difference in texture could explain the slope difference in the case of surfaces lacking regolith.

- Space weathering induces alteration of the minerals on the surface of asteroids and is intensely studied in the laboratory as its effects strongly depend on the initial composition of the sample and the dose of irradiation. Reddening, as well as bluing, of the spectra have been observed after laboratory irradiation of meteorites (Brunetto et al., 2014; Lantz et al., 2018, 2015; Matsuoka et al., 2015; Moroz et al., 2002; Strazzulla et al., 2005). Lantz et al. (2018) found that space weathering on CI- or CM chondrite-like material can induce a blue slope on their reflectance spectra, with more than 60% of variation for the highest irradiation fluences (Lantz et al., 2017). Kaluna et al. (2017) explained this blue slope by the production of large organic-like particles. Space weathering can thus be an explanation for the spectral slope difference between the irradiated surfaces of asteroids and our samples protected from any irradiation during the experiment.

c. The missing organics

Our experiments have shown that a 90 minutes-long heating at 523 K induces an appearance or increase of the amplitude of the 3.4-3.5 µm organic bands. These organic features should thus be clearly detectable on reflectance spectra of asteroids, especially NEAs, taking into account their age and surface temperature.

However, so far, organics features were never detected among C-complex asteroids, with the exception of local signatures detected at the surface of Ceres (DeSanctis et al., 2017). Given the high-quality data available on the largest objects as well as on objects visited by space missions, if present at the level found in meteorites, organics should have been observed in asteroid spectra. Long-term heating experiments on the insoluble organic matter and bulk of the meteorite Murchison have shown an instability of the aliphatic C-H bounds, so CH2 and CH3. According to calculations by Kebukawa et al. (2010), all aliphatic bounds are lost in approximately 200 years if placed at 373K, and in 90s at 773K. This instability could explain the lack of organic signatures in the reflectance spectra of heated small bodies such as NEAs. Several other possibilities can explain this lack of organic signatures in the reflectance spectra of MBAs.

A first explanation is that carbonaceous chondrites are not representative of C-type asteroids. This is a strong possibility for the least hydrated C-types (Vernazza and Beck, 2017) but more difficult to imagine the case of Ch- and Cgh-types. Indeed, there are three spectral features (0.7, 0.9 and 3 µm) that are seen in both these asteroids types and CM chondrites, which imposes a strong connection.

It has also been shown that organics signatures ($-CH_2$ and $–CH_3$) can be destroyed by a space weathering process. Lantz et al. (2015) studied the effect of irradiation with high-energy ions on carbonaceous chondrites, simulating the effect of solar wind and cosmic rays on the surface of asteroids. They showed that irradiation by the solar wind leads to an amorphization of the poly-aromatic structure of the organics. The amorphization process may induce a loss of hydrogen and consequently a change in the infrared signatures at 3.4-3.5 µm. However, Lantz et al. (2017) showed that solar wind irradiation should also destroy the 3-µm band. A disappearance of the organics feature should thus be associated to a disappearance of the 3-µm band.

If solar wind weathering fails to explain the disappearance of the organic feature, other type of space weathering might affect the surface optical properties. Galactic Cosmic Rays could play a role in the destruction of the organic signatures, as well as micrometeorites impacts. Still, these two processes are also expected to affect both the 3-µm band and the organic signatures.

A last and favored possibility is the impact of intense UV irradiation from the Sun. This process offers an interesting possibility since organic compounds tend to have high absorption coefficients in the UV, while water molecules will be less affected. When exposed to UV radiation, the loss of hydrogen and carbon (in the form of $CH_4$) has been shown in the case of the Murchison meteorite (Keppler et al., 2012), which could result in a decrease of the organic features at 3.4-3.5 µm. Still, this mechanism needs to be validated by laboratory experiments.

5. Conclusions

Our results show that an asteroid-like environment induces irreversible alterations to carbonaceous chondrites. Because of the dehydration of the sample, the 3-µm absorption band becomes shallower, thinner and sharper. Interestingly, the most aqueously altered chondrites show the smallest spectral variations, and this can be explained by their high abundance of phyllosilicates that are unaltered at temperatures up to 523 K. We detected an increase of the amplitude of the organics features around 3.4 µm and 3.5 µm on the heated samples.

The 3-µm absorption bands detected by AKARI on MBAs are coherent with those observed on our heated samples. All bands are centered around 2.75 µm, but the band depths are generally weaker and less dispersed than the meteorites. This difference can be explained by the irradiation of the surface of the small bodies as well as the difference of texture between the asteroids and our powdered samples. Moreover, the band is thinner and presents a more symmetrical shape on the asteroids spectra compared to the bands detected on the meteorites. The duration of the heating in our experiment can be the cause of such a difference, and further analyses on the effects of long-term heating are needed. We also highlighted that the asteroids spectra are generally bluer than the meteorites spectra, though this could be due to an effect of the space weathering of the small bodies surfaces, and a difference of surface texture between our powdered chondrites and possibly lacking regolith asteroids.

The deconvolution of the 3-µm band in both meteorites and asteroids spectra highlighted the effect of aqueous and thermal alteration on the surfaces. Consistency between the components reaffirms the link between the CM, CR and CI carbonaceous chondrites and C-complex asteroids. As for meteorites, a metal-OH component is always detected in the 3-µm band of asteroids, and most of them show signs of water molecules. However, the components detected on the asteroids spectra tend to be weaker than on our meteorites measurements. It has been observed that asteroids from the same group present roughly the same components. Focusing on the NEAs, the 3-µm band of Ryugu shows the scars of a heavy aqueous alteration followed by a heating episode, while Bennu presents also an aqueously altered surface as well as hints of its surface activity through the potential presence of adsorbed water. We also found that the asteroid Dione, found to be active, also displays adsorbed water and it is anticipated that the other Cgh-type asteroid studied, Nemausa, may be also active.

Acknowledgements:
This research is based on observation with AKARI, a JAXA project with the participation of ESA. SP is supported by Université Grenoble Alpes (UGA) (IRS IDEX/UGA). PB acknowledges funding from the European Research Council under the grant SOLARYS ERC-CoG2017-771691. FU is supported by JSPS KAKENHI (Grants-in-Aid for Scientific Research, Grants N° JP17K05636 and JP19H00725). The instrument SHADOWS was founded by the OSUG@2020 Labex (Grant ANR10 LABX56), by 'Europlanet 2020 RI' within the European Union's Horizon 2020 research and innovation program (Grant N° 654208) and by the Centre National d'Etudes Spatiales (CNES).

APPENDIX A: Modeled spectra of meteorites

The modeled spectra of the meteorites are presented in this section, before heating at 523 K on the left, after on the right.

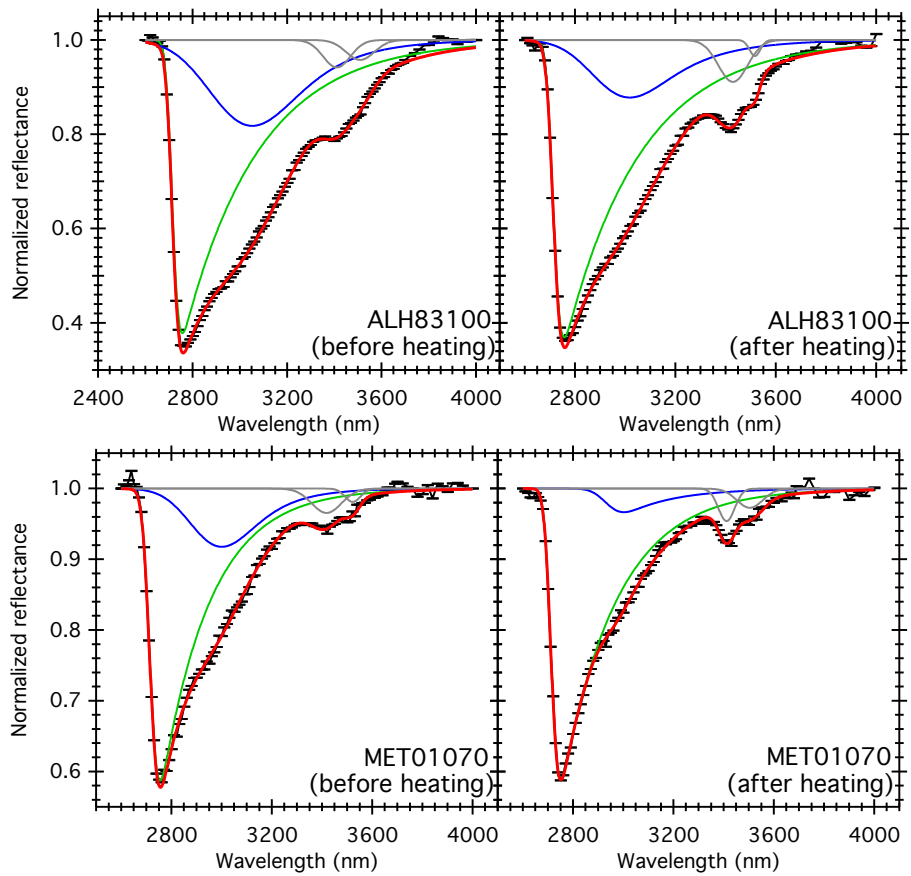

Figure 11: Modeled spectra of the CM1 chondrites. Black: reflectance measurement, Red: modeled spectrum, Green: metal-OH component, Blue: structural/interlayer water molecules component. Grey: organic bands.

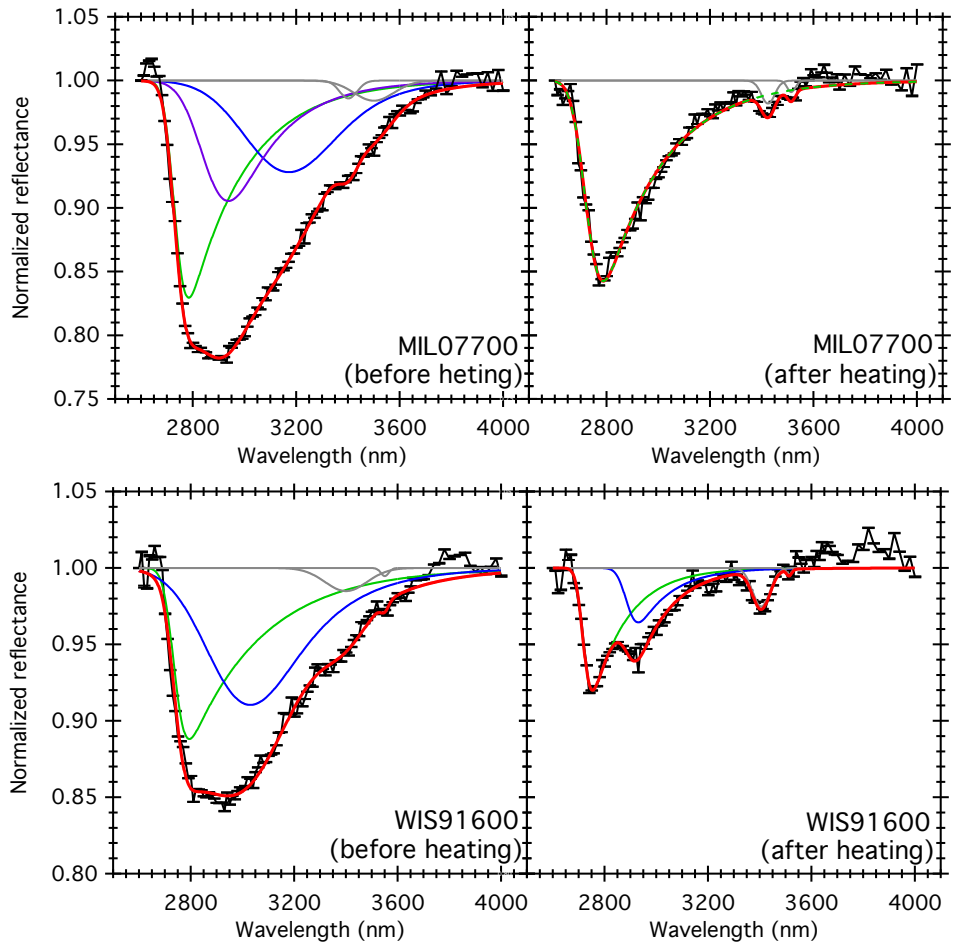

Figure 12: Modeled spectra of the CM2 chondrites. Black: reflectance measurement, Red: modeled spectrum, Green: metal-OH component, Blue: structural/interlayer water molecules component. Grey: organic bands.

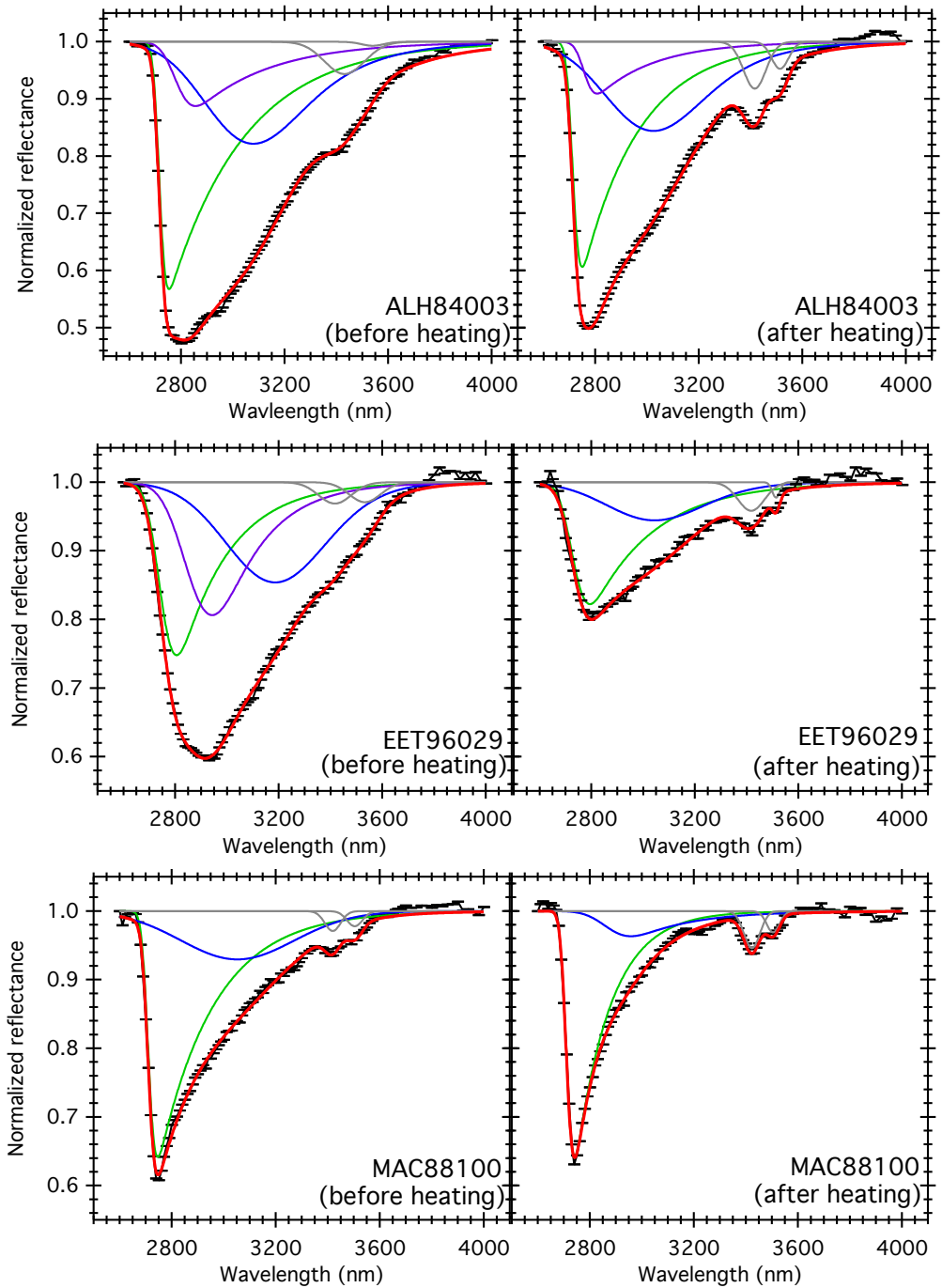

Figure 13: Modeled spectra of the heated CM2 chondrites. Black: reflectance measurement, Red: modeled spectrum, Green: metal-OH component, Blue: structural/interlayer water molecules component. Grey: organic bands.

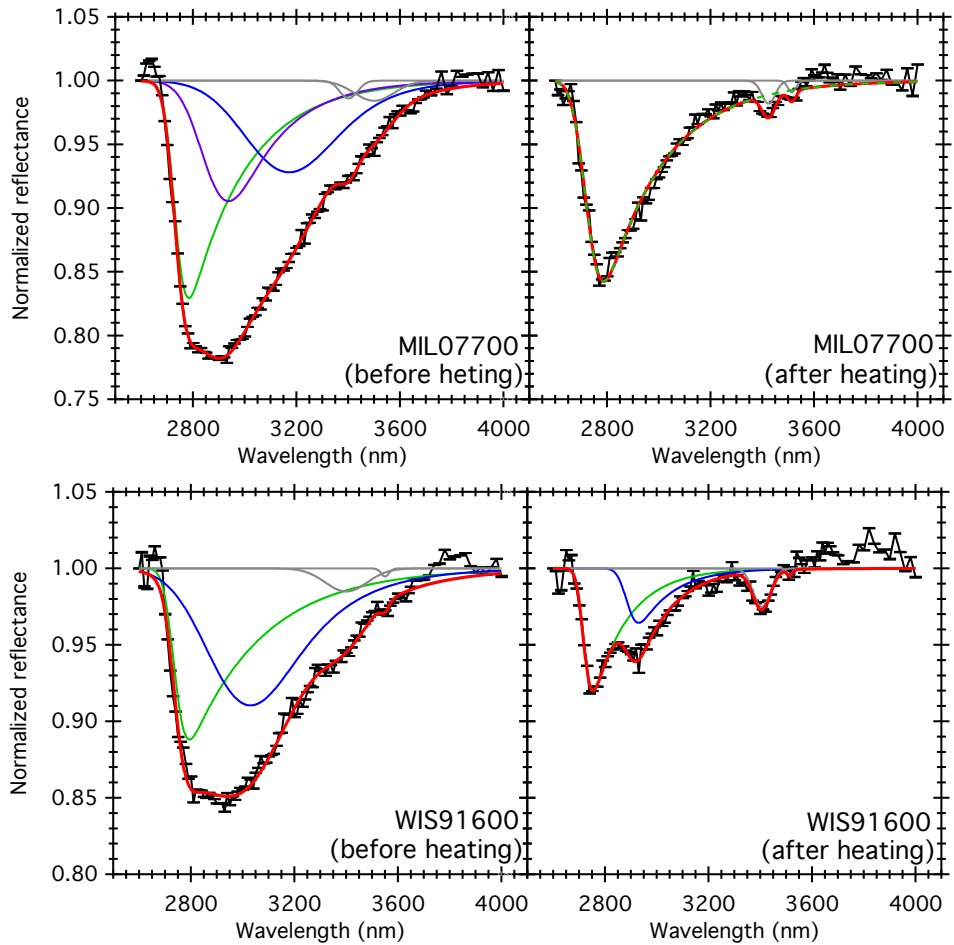

Figure 15 (end): Modeled spectra of the heated CM2 chondrites. Black: reflectance measurement, Red: modeled spectrum, Green: metal-OH component, Blue: structural/interlayer water molecules component. Grey: organic bands.

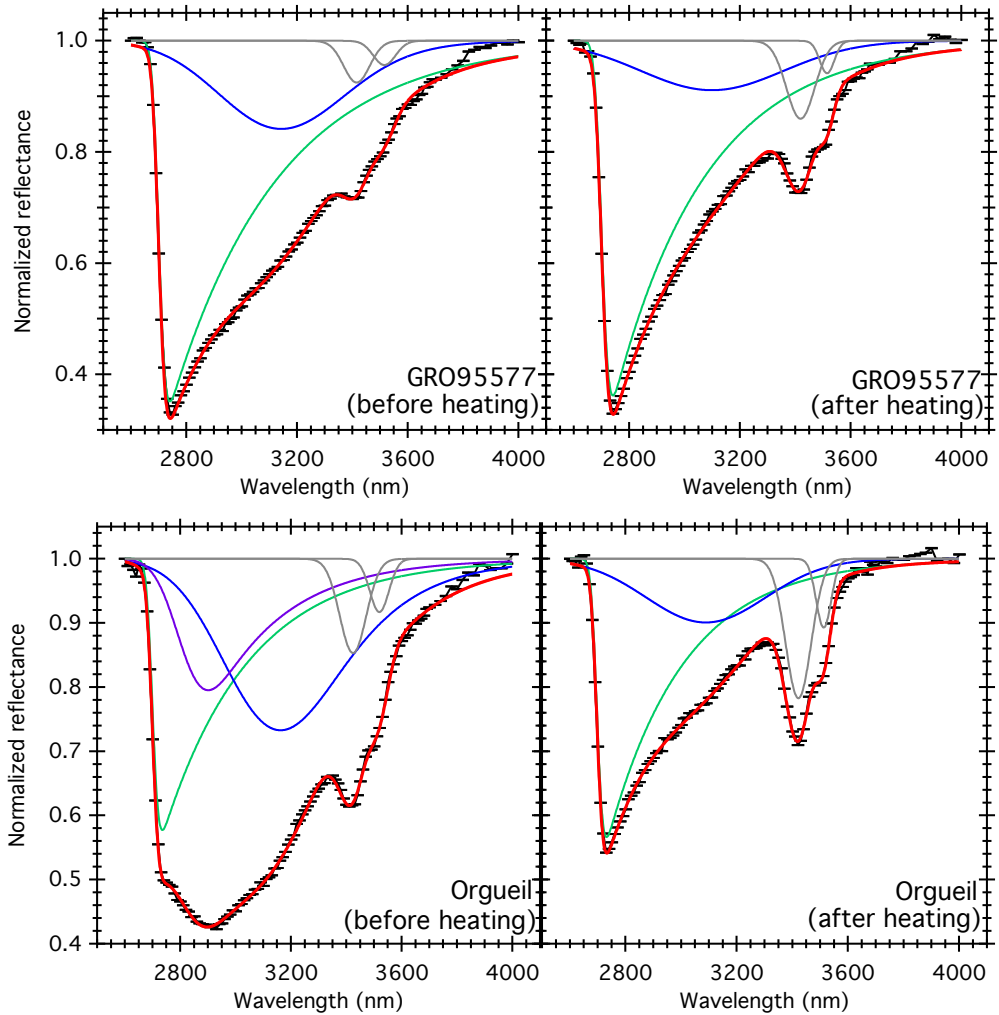

Figure 14: Modeled spectra of the CR1 chondrite GRO 95577 and the CI1 Orgueil. Black: reflectance measurement, Red: modeled spectrum, Green: metal-OH component, Blue: structural/interlayer water molecules component. Grey: organic bands.

APPENDIX B: Modeled spectra of asteroids
The modeled spectra of the asteroids are presented in this section.

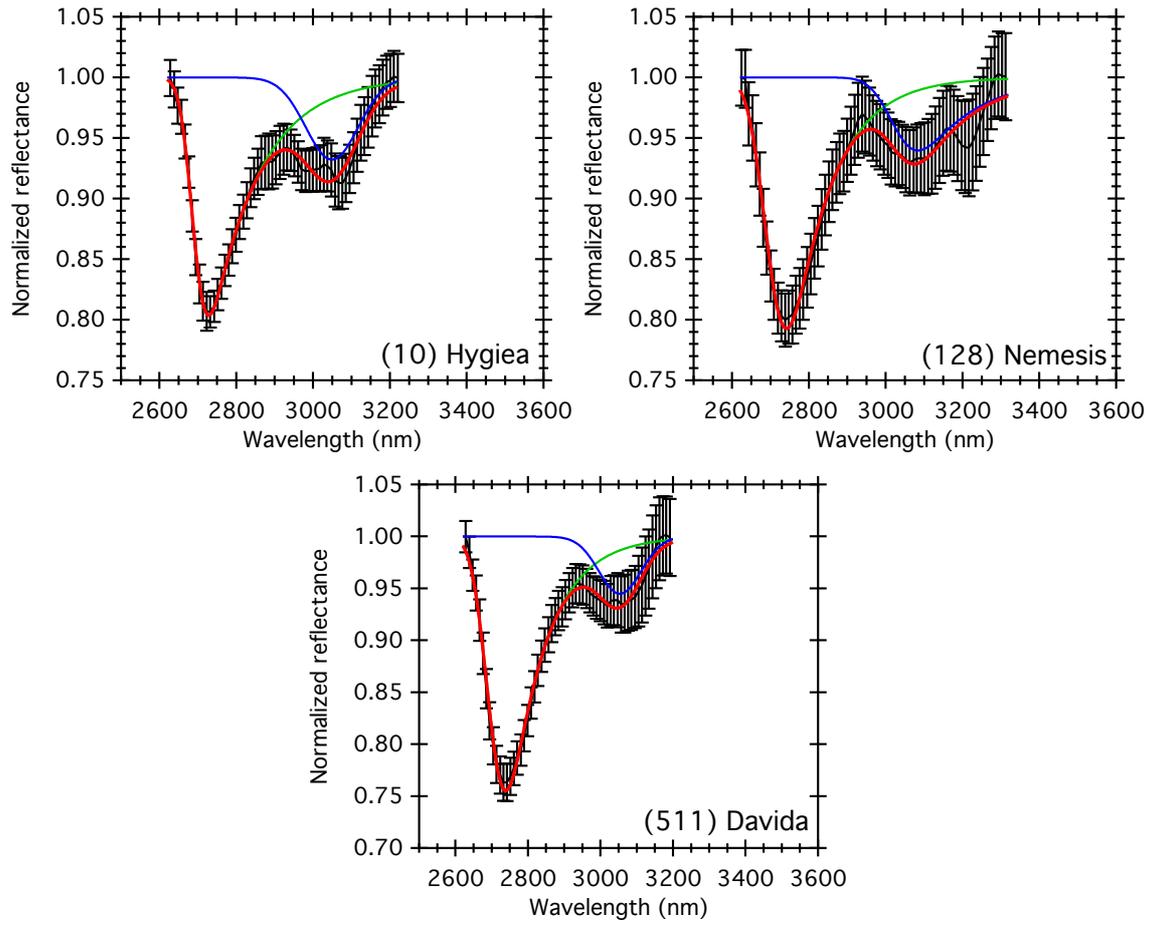

Figure 15: Modeled spectra of the C-type asteroids. Black: observation data, Red: modeled spectrum, Green: metal-OH component, Blue: water molecules component.

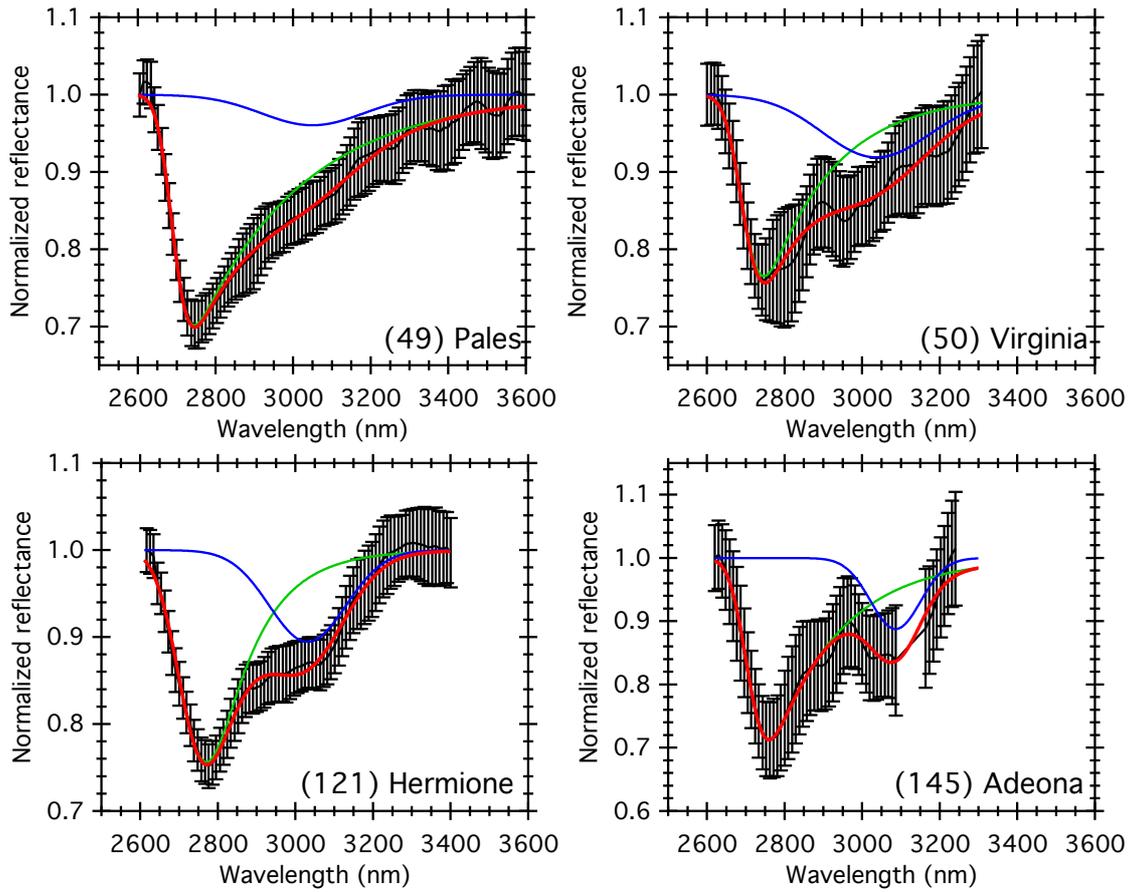

Figure 16: Modeled spectra of the Ch-type asteroids. Black: observation data, Red: modeled spectrum, Green: metal-OH component, Blue: water molecules component.

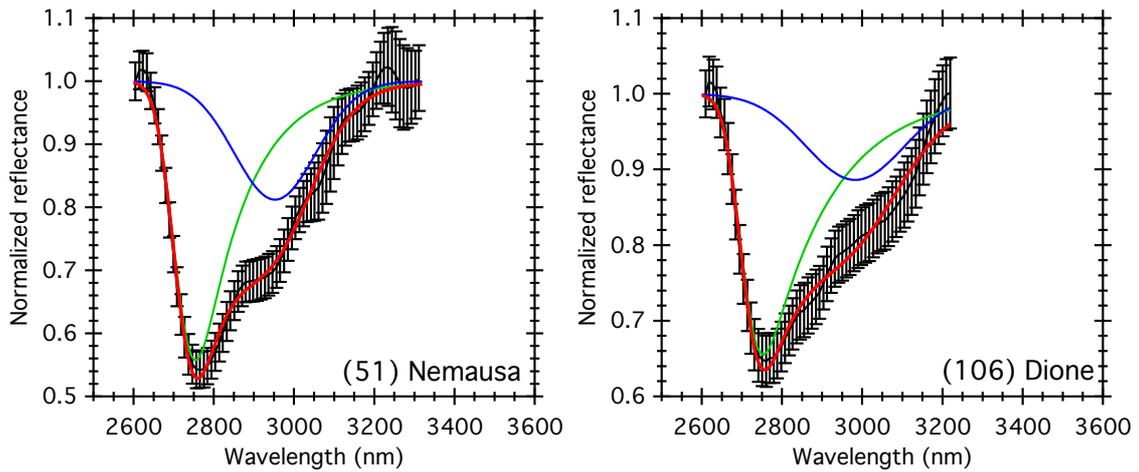

Figure 20: Modeled spectra of the Cgh-type asteroids. Black: observation data, Red: modeled spectrum, Green: metal-OH component, Blue: water molecules component.

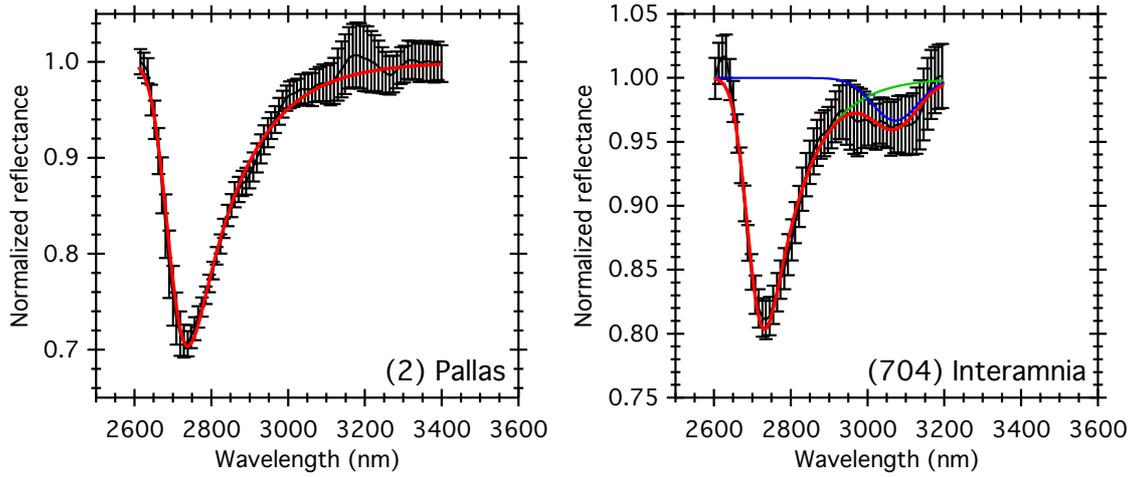

Figure 17: Modeled spectra of the B-type asteroid (2) Pallas and Cb-type (704) Interamnia. Black: observation data, Red: modeled spectrum, Green: metal-OH component, Blue: water molecules component.

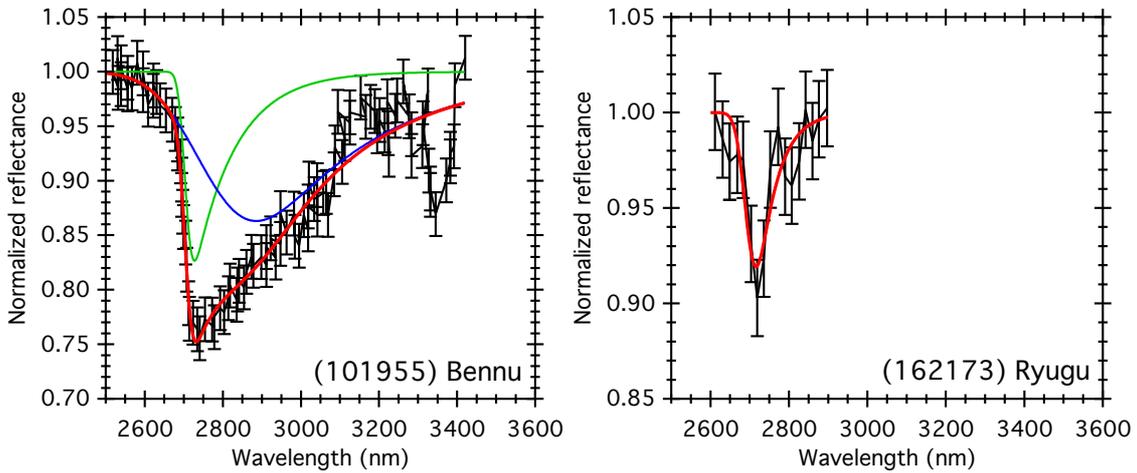

Figure 18: Modeled spectra of the NEAs. Black: observation data, Red: modeled spectrum, Green: metal-OH component, Blue: water molecules component.

APPENDIX C: Parameters of the metal-OH, HFW and LFW components of asteroids and meteorites derived from the modeled spectra.

| Asteroids | Type | | Complete band | Metal-OH | LFW | HFW |
|---|---|---|---|---|---|---|
| (2) Pallas | B | Band depth (%) | 29.7 ± 0.8 | 29.7 ± 0.8 | - | - |
| | | Position (nm) | 2738 ± 4 | 2738 ± 4 | - | - |
| | | FWHM (nm) | 175 ± 5 | 175 ± 5 | - | - |
| | | Symmetry factor | 0.511 ± 0.03 | 0.511 ± 0.03 | - | - |
| (10) Hygeia | C | Band depth (%) | 19.70 ± 0.67 | 19.7 ± 0.7 | 7.0 ± 0.9 | - |
| | | Position (nm) | 2727.9 ± 4.9 | 2728 ± 5 | 3049 ± 10 | - |
| | | FWHM (nnm) | 149.7 ± 9.5 | 149 ± 11 | 158 ± 24 | - |
| | | Symmetry factor | 0.455 ± 0.070 | 0.457 ± 0.07 | 0.996 ± 0.005 | - |
| (49) Pales | Ch | Band depth (%) | 29.9 ± 1.3 | 29.0 ± 1.9 | 7.3 ± 2.4 | - |
| | | Position (nm) | 2746 ± 8 | 2744 ± 9 | 3001 ± 41 | - |

| Asteroid | Type | Parameter | | | | |
|---|---|---|---|---|---|---|
| | | FWHM (nnm) | 308 ± 34 | 203 ± 27 | 304 ± 40 | - |
| | | Symmetry factor | 0.25 ± 0.04 | 0.413 ± 0.09 | 0.993 ± 0.009 | - |
| (50) Virginia | Ch | Band depth (%) | 24.7 ± 2.6 | 24.1 ± 3.2 | 9.7 ± 4.7 | - |
| | | Position (nm) | 2749 ± 20 | 2748 ± 19 | 2998 ± 64 | - |
| | | FWHM (nnm) | 380 ± 51 | 207 ± 68 | 265 ± 162 | - |
| | | Symmetry factor | 0.199 ± 0.090 | 0.994 ± 0.01 | 0.963 ± 0.091 | - |
| (51) Nemausa | Cgh | Band depth (%) | 47.6 ± 1.3 | 45.8 ± 2.6 | - | 18.2 ± 3.1 |
| | | Position (nm) | 2760.5 ± 5.7 | 2756 ± 7 | - | 2951 ± 14 |
| | | FWHM (nnm) | 300 ± 11 | 171 ± 13 | - | 219 ± 30 |
| | | Symmetry factor | 0.29 ± 0.03 | 0.600 ± 0.09 | - | 0.9947 ± 0.0052 |
| (106) Dione | Cgh | Band depth (%) | 37.0 ± 1.5 | 34.6 ± 3.0 | - | 14.2 ± 2.2 |
| | | Position (nm) | 2758.6 ± 9.9 | 2753 ± 11 | - | 2964 ± 30 |
| | | FWHM (nnm) | 311 ± 17 | 171 ± 16 | - | 270 ± 35 |
| | | Symmetry factor | 0.278 ± 0.048 | 0.614 ± 0.07 | - | 0.9955 ± 0.0033 |
| (121) Hermione | Ch | Band depth (%) | 24.60 ± 0.94 | 24.5 ± 1.0 | 10.0 ± 1.6 | - |
| | | Position (nm) | 2770.8 ± 7.7 | 2769 ± 8 | 3035 ± 19 | - |
| | | FWHM (nnm) | 379 ± 19 | 203 ± 18 | 216 ± 36 | - |
| | | Symmetry factor | 0.27 ± 0.03 | 0.658 ± 0.09 | 0.994 ± 0.007 | - |
| (128) Nemesis | C | Band depth (%) | 20.84 ± 0.99 | 20.8 ± 1.0 | 6.4 ± 1.1 | - |
| | | Position (nm) | 271.90 ± 8.4 | 2742 ± 8 | 3081 ± 50 | - |
| | | FWHM (nnm) | 161 ± 10 | 162 ± 11 | 201 ± 47 | - |
| | | Symmetry factor | 0.63 ± 0.12 | 0.632 ± 0.13 | 0.993 ± 0.010 | - |
| (145) Adeona | Ch | Band depth (%) | 29.4 ± 3.0 | 29.3 ± 3.0 | 13.4 ± 4.3 | - |
| | | Position (nm) | 2640 ± 14 | 2760 ± 14 | 3074 ± 51 | - |
| | | FWHM (nnm) | 89 ± 59 | 198 ± 40 | 137 ± 62 | - |
| | | Symmetry factor | 0.29 ± 0.52 | 0.474 ± 0.17 | 0.996 ± 0.010 | - |
| (511) Davida | C | Band depth (%) | 24.46 ± 0.79 | 24.5 ± 0.8 | 5.7 ± 1.2 | - |
| | | Position (nm) | 2738.1 ± 4.7 | 2738 ± 5 | 3051 ± 16 | - |
| | | FWHM (nnm) | 151.9 ± 6.6 | 152 ± 7 | 136 ± 32 | - |
| | | Symmetry factor | 0.623 ± 0.073 | 0.622 ± 0.08 | 0.993 ± 0.010 | - |
| (704) Interamnia | Cb | Band depth (%) | 20.53 ± 0.69 | 20.5 ± 0.7 | 3.9 ± 0.9 | - |
| | | Position (nm) | 2732.1 ± 4.6 | 2732 ± 5 | 3069 ± 24 | - |
| | | FWHM (nnm) | 146.1 ± 7.5 | 146 ± 8 | 130 ± 41 | - |
| | | Symmetry factor | 0.607 ± 0.076 | 0.606 ± 0.08 | 0.992 ± 0.011 | - |
| (101955) Bennu | NEA | Band depth (%) | 24.3 ± 1.3 | 21.4 ± 3.9 | - | 11.5 ± 2.1 |
| | | Position (nm) | 2740 ± 13 | 2737 ± 13 | - | 2909 ± 39 |
| | | FWHM (nnm) | 261 ± 13 | 110 ± 24 | - | 251 ± 47 |
| | | Symmetry factor | 0.190 ± 0.061 | 0.534 ± 0.11 | - | 0.9966 ± 0.0004 |
| (162173) Ryugu | NEA | Band depth (%) | 8.4 ± 2.5 | 8.4 ± 2.5 | - | - |
| | | Position (nm) | 2715 ± 8 | 2715 ± 8 | - | - |
| | | FWHM (nnm) | 62 ± 62 | 62 ± 62 | - | - |
| | | Symmetry factor | 0.701 ± 0.31 | 0.701 ± 0.31 | - | - |

Table 3: Parameters of the metal-OH, HFW and LFW components derived from the asteroid spectra. The minus sign indicates that the component is not detected.

| Meteorites | Type | | Complete band | | Metal-OH | | LFW | | HFW | |
|---|---|---|---|---|---|---|---|---|---|---|
| | | | before | after | before | after | before | after | before | after |
| ALH 83100 | CM1 | Band depth (%) | 64.14 ± 0.59 | 62.37 ± 0.27 | 62.448 ± 0.008 | 63.048 ± 0.012 | 17.786 ± 0.008 | 12.200 ± 0.015 | - | - |
| | | Position (nm) | 2770 ± 5 | 2760 ± 5 | 2756.95 ± 0.02 | 2756.96 ± 0.03 | 3052.77 ± 0.06 | 3018.05 ± 0.08 | - | - |
| | | FWHM (nm) | 470.0 ± 7.1 | 390.0 ± 7.1 | 277.04 ± 0.07 | 267.25 ± 0.12 | 451.15 ± 0.11 | 395.97 ± 0.30 | - | - |
| | | Symmetry factor | 0.146 ± 0.012 | 0.147 ± 0.015 | 0.35647 ± 0.01213 | 0.21277 ± 0.00014 | 0.64511 ± 0.02078 | 0.75728 ± 0.00062 | - | - |
| MET 01070 | CM1 | Band depth (%) | 43.994 ± 0.95 | 40.45 ± 0.36 | 41.281 ± 0.009 | 41.155 ± 0.005 | 8.230 ± 0.013 | 3.332 ± 0.011 | - | - |
| | | Position (nm) | 2760 ± 5 | 2750 ± 5 | 2754.67 ± 0.03 | 2751.41 ± 0.03 | 3000.17 ± 0.08 | 3002.00 ± 0.21 | - | - |
| | | FWHM (nnm) | 270.0 ± 7.1 | 230 ± 7.7 | 193.29 ± 0.11 | 206.22 ± 0.12 | 320.06 ± 0.24 | 244.86 ± 0.72 | - | - |
| | | Symmetry factor | 0.227 ± 0.023 | 0.211 ± 0.026 | 0.30668 ± 0.00031 | 0.26352 ± 0.00024 | 0.81029 ± 0.00083 | 0.49185 ± 0.00177 | - | - |
| DOM 08003 | CM2 | Band depth (%) | 64.14 ± 0.59 | 55.04 ± 0.44 | 48.123 ± 0.039 | 47.907 ± 0.106 | 25.512 ± 0.017 | 23.834 ± 0.036 | 10.950 ± 0.033 | 8.753 ± 0.086 |
| | | Position (nm) | 2770 ± 5 | 2770 ± 5 | 2745.16 ± 0.04 | 2743.50 ± 0.08 | 3069.73 ± 0.07 | 304387 ± 0.09 | 2822.58 ± 0.19 | 2819.00 ± 0.31 |
| | | FWHM (nnm) | 470.0 ± 7.1 | 460.0 ± 7.1 | 256.59 ± 0.16 | 197.98 ± 0.25 | 479.85 ± 0.14 | 453.54 ± 0.19 | 293.33 ± 0.19 | 225.83± 0.61 |
| | | Symmetry factor | 0.146 ± 0.012 | 0.150 ± 0.013 | 0.18891 ± 0.00020 | 0.24667 ± 0.00048 | 0.75325 ± 0.00031 | 0.76749 ± 0.00044 | 0.30938 ± 0.00040 | 0.37083 ± 0.00140 |
| Murchison | CM2 | Band depth (%) | 35.9 ± 0.32 | 33.57 ± 0.47 | 32.411 ± 0.010 | 32.087 ± 0.010 | 14.238 ± 0.009 | 10.212 ± 0.012 | - | - |
| | | Position (nm) | 2820 ± 5 | 2810 ± 5 | 2799.91 ± 0.04 | 2794.53 ± 0.03 | 3076.00 ± 0.10 | 3063.38 ± 0.09 | - | - |
| | | FWHM (nnm) | 505.0 ± 7.1 | 440.0 ± 7.1 | 310.88 ± 0.12 | 266.23 ± 0.15 | 486.93 ± 0.17 | 411.75 ± 0.20 | - | - |
| | | Symmetry factor | 0.188 ± 0.012 | 0.257 ± 0.015 | 0.28430 ± 0.00022 | 0.35438 ± 0.00033 | 0.83760 ± 0.00048 | 0.72917 ± 0.00098 | - | - |
| QUE 97990 | CM2 | Band depth (%) | 35.62 ± 0.39 | 29.69 ± 0.42 | 33.294 ± 0.018 | 28.814 ± 0.007 | 10.399 ± 0.016 | 8.120 ± 0.008 | 12.190 ± 0.015 | - |
| | | Position (nm) | 2830 ± 5 | 2820 ± 5 | 2805.44 ± 0.04 | 2808.23 ± 0.04 | 3192.05 ± 0.24 | 3063.35 ± 0.15 | 2968.89 ± 0.13 | - |
| | | FWHM (nnm) | 52.0 ± 7.1 | 420.0 ± 7.1 | 243.17 ± 0.13 | 280.73 ± 0.11 | 439.61 ± 0.27 | 433.66 ± 0.17 | 302.38 ± 0.34 | - |
| | | Symmetry factor | 0.209 ± 0.012 | 0.235 ± 0.015 | 0.35665 ± 0.00035 | 0.30342 ± 0.00026 | 0.83892 ± 0.00068 | 0.83031 ± 0.00068 | 0.59697 ± 0.00081 | - |
| ALH84 003 | CM2 (h) | Band depth (%) | 51.31 ± 0.31 | 48.82 ± 0.31 | 43.267 ± 0.018 | 39.364 ± 0.096 | 17.833 ± 0.016 | 15.592 ± 0.012 | 11.239 ± 0.015 | 9.068 ± 0.088 |
| | | Position (nm) | 2800 ± 5 | 2770 ± 5 | 2753.21 ± 0.03 | 2749.15 ± 0.07 | 3078.80 ± 0.22 | 3024.41 ± 0.10 | 2855.25 ± 0.34 | 2806.63 ± 0.25 |
| | | FWHM (nnm) | 500.0 ± 7.1 | 370.0 ± 7.1 | 252.61 ± 0.11 | 204.53 ± 0.11 | 479.87 ± 0.16 | 467.82 ± 0.15 | 311.24 ± 0.25 | 225.74 ± 0.27 |
| | | Symmetry factor | 0.190 ± 0.012 | 0.194 ± 0.016 | 0.18978 ± 0.00017 | 0.24231 ± 0.00031 | 0.80938 ± 0.00038 | 0.85190 ± 0.00041 | 0.38468 ± 0.00121 | 0.32679 ± 0.00081 |
| EET 96029 | CM2 (h) | Band depth (%) | 39.70 ± 0.27 | 19.44 ± 0.42 | 25.207 ± 0.019 | 17.7638 ± 0.008 | 14.591 ± 0.015 | 5.559 ± 0.008 | 19.369 ± 0.014 | - |
| | | Position (nm) | 2930 ± 5 | 2790 ± 5 | 2805.17 ± 0.06 | 2794.13 ± 0.07 | 3186.17 ± 0.24 | 3043.76 ± 0.26 | 2942.39 ± 0.13 | - |
| | | FWHM (nnm) | 440.0 ± 7.1 | 340.0 ± 7.1 | 233.93 ± 0.14 | 268.68 ± 0.14 | 457.58 ± 0.24 | 444.93 ± 0.21 | 299.28 ± 0.16 | - |
| | | Symmetry factor | 0.692 ± 0.019 | 0.308 ± 0.019 | 0.47134 ± 0.00049 | 039943 ± 0.00044 | 0.95073 ± 0.00032 | 0.93515 ± 0.00065 | 0.67837 ± 0.00052 | - |
| MAC 88100 | CM2 (h) | Band depth (%) | 39.63 ± 0.28 | 37.87 ± 0.42 | 35.833 ± 0.009 | 35.742 ± 0.006 | 7.035 ± 0.012 | 3.660 ± 0.025 | - | - |
| | | Position (nm) | 2750 ± 5 | 2740 ± 5 | 2746.30 ± 0.05 | 2741.50 ± 0.02 | 3052.20 ± 0.17 | 2957.46 ± 0.37 | - | - |
| | | FWHM (nnm) | 230.0 ± 7.1 | 140.0 ± 7.1 | 197.72 ± 0.13 | 138.06 ± 0.19 | 514.68 ± 0.36 | 307.46 ± 0.46 | - | - |
| | | Symmetry factor | 0.210 ± 0.026 | 0.273 ±45 | 0.25249 ± 0.00030 | 0.34602 ± 0.00064 | 0.98515 ± 0.00023 | 0.56267 ± 0.00104 | - | - |
| MIL 07700 | CM2 (h) | Band depth (%) | 21.02 ± 0.75 | 15.34 ± 0.39 | 17.039 ± 0.013 | 15.748 ± 0.004 | 7.192 ± 0.012 | - | 9.457 ± 0.010 | - |
| | | Position (nm) | 2930 ± 5 | 2780 ± 5 | 2784.10 ± 0.06 | 2788.28 ± 0.06 | 3171.43 ± 0.28 | - | 2937.59 ± 0.19 | - |
| | | FWHM (nnm) | 455.0 ± 7.1 | 260.0 ± 7.1 | 234.49 ± 0.22 | 260.99 ± 0.08 | 427.12 ± 0.30 | - | 301.26 ± 0.23 | - |
| | | Symmetry factor | 0.717 ± 0.019 | 0.316 ± 0.026 | 0.35899 ± 0.00051 | 0.43426 ± 0.00041 | 0.86839 ± 0.00062 | - | 0.64158 ± 0.00077 | - |
| WIS 91600 | CM2 (h) | Band depth (%) | 14.97 ± 0.49 | 6.92 ± 0.81 | 11.186 ± 0.009 | 8.001 ± 0.005 | 8.953 ± 0.009 | 3.555 ± 0.008 | - | - |
| | | Position (nm) | 2930 ± 5 | 2740 ± 5 | 2794.32 ± 0.22 | 2753.43 ± 0.07 | 3029.91 ± 0.23 | 2930.00 ± 0.14 | - | - |
| | | FWHM (nnm) | 390.0 ± 7.1 | 220.0 ± 7.1 | 280.17 ± 0.20 | 148.22 ± 0.29 | 440.20 ± 0.18 | 162.42 ± 0.44 | - | - |
| | | Symmetry factor | 0.950 ± 0.025 | 0.157 ± 0.026 | 0.30053 ± 0.00092 | 0.38351 ± 0.00112 | 0.77655 ± 0.00083 | 0.47944 ± 0.00167 | - | - |
| GRO 95577 | CR1 | Band depth (%) | 66.64 ± 0.21 | 65.56 ± 0.20 | 64.857 ± 0.007 | 63.781 ± 0.008 | 15.846 ± 0.008 | 8.881 ± 0.008 | - | - |
| | | Position (nm) | 2740 ± 5 | 2740 ± 5 | 2740.67 ± 0.02 | 2741.156 ± 0.05 | 3143.04 ± 0.08 | 3098.81 ± 0.16 | - | - |
| | | FWHM (nnm) | 490.0 ± 7.1 | 420.0 ± 7.1 | 324.09 ± 0.08 | 284.91 ± 0.07 | 542.16 ± 0.16 | 609.49 ± 0.32 | - | - |
| | | Symmetry factor | 0.113 ± 0.011 | 0.105 ± 0.013 | 0.15011 ± 0.00004 | 0.17816 ± 0.00024 | 0.94418 ± 0.00031 | 0.99130 ± 0.00016 | - | - |
| Orgueil | CI1 | Band depth (%) | 56.51 ± 0.53 | 46.55 ± 0.49 | 45.864 ± 0.004 | 42.264 ± 0.018 | 26.717 ± 0.020 | 9.926 ± 0.008 | 20.473 ± 0.018 | - |
| | | Position (nm) | 2930 ± 5 | 2730 ± 5 | 2732.87 ± 0.03 | 2735.56 ± 0.03 | 3162.43 ± 0.11 | 3088.52 ± 0.09 | 2901.05± 0.09 | - |
| | | FWHM (nnm) | 680.0 ± 7.1 | 350.0 ± 7.1 | 374.94 ± 0.08 | 261.26 ± 0.19 | 528.96 ± 0.23 | 506.53 ± 0.24 | 350.18 ± 0.15 | - |
| | | Symmetry factor | 0.511 ± 0.011 | 0.166 ± 0.017 | 0.11341 ± 0.00014 | 0.16914 ± 0.00019 | 0.81571 ± 0.00034 | 0.99200 ± 0.00005 | 0.52994 ± 0.00040 | - |

Table 4: Parameters of the metal-OH, HFW and LFW components derived from the meteorite spectra. The minus sign indicates that the component is not detected.